\documentclass[usenatbib,usegraphicx,useAMS]{mn2e}
\usepackage{lscape}
\usepackage{multirow}
\usepackage{comment}
\usepackage{longtable}

\newcommand\aj{{AJ}}%
\newcommand\apj{{ApJ}}%
\newcommand\apjs{{ApJS}}%
\newcommand\aap{{A\&A}}%
\newcommand\aaps{{A\&AS}}%
\newcommand\mnras{{MNRAS}}%
\newcommand\pasp{{PASP}}%
\newcommand{\NUMpleiads}{\ensuremath{470}}
\newcommand{\NUMpleiadstot}{\ensuremath{488}}
\newcommand{\NUMpleiadsnotinstauffer}{\ensuremath{18}}
\newcommand{\NUMpleiadsnew}{\ensuremath{14}}
\newcommand{\NUMperpleiads}{\ensuremath{350}}

\newcommand{\NUMperpleiadsfoundbyhat}{\ensuremath{368}}
\newcommand{\NUMperprevknown}{\ensuremath{66}}
\newcommand{\NUMperprevknownperfoundmatch}{\ensuremath{50}}

\newcommand{\NUMperprevknowntoofaint}{\ensuremath{9}}
\newcommand{\NUMperprevknownperfoundmatchclosefreq}{\ensuremath{44}}
\newcommand{\NUMperprevknownperfoundmatchrealclosefreq}{\ensuremath{35}}
\newcommand{\NUMwithperandspec}{\ensuremath{223}}

\title[Pleiades Rotation Periods]{A Large Sample of Photometric Rotation Periods for FGK Pleiades Stars}
\author[J. D. Hartman et al.]{J. D. Hartman$^{1}$\thanks{E-mail: jhartman@cfa.harvard.edu (JDH)}, G. \'A. Bakos$^{1}$, G. Kov\'acs$^{2}$, R. W. Noyes$^{1}$\\
$^{1}$Harvard-Smithsonian Center for Astrophysics, 60 Garden St., Cambridge, MA~02138, USA\\
$^{2}$Konkoly Observatory, Budapest, Hungary.\\
}
\begin{document}

\pagerange{\pageref{firstpage}--\pageref{lastpage}} \pubyear{2010}

\maketitle

\label{firstpage}

\begin{abstract}
Using data from the HATNet survey for transiting exoplanets we measure
photometric rotation periods for \NUMperpleiadsfoundbyhat{} Pleiades
stars with $0.4~M_{\odot} \la M \la 1.3~M_{\odot}$. We detect periodic
variability for 74\% of the cluster members in this mass range that
are within our field-of-view, and 93\% of the members with
$0.7~M_{\odot} \la M \la 1.0~M_{\odot}$. This increases, by a factor
of five, the number of Pleiades members with measured
periods. Included in our sample are \NUMpleiadsnew{} newly identified
probable cluster members which have proper motions, photometry, and
rotation periods consistent with membership. We compare this data to
the rich sample of spectroscopically determined projected equatorial
rotation velocities ($v \sin i$) available in the literature for this
cluster. For stars with $M \ga 0.85~M_{\odot}$ the rotation periods,
$v \sin i$ and radius estimates are consistent with the stars having
an isotropic distribution of rotation axes, if a moderate differential
rotation law is assumed. For stars with $M \la 0.85~M_{\odot}$ the
inferred $\sin i$ values are systematically larger than $1.0$. These
observations imply that the combination of measured parameters $P (v
\sin i) / R$ is too large by $\sim 24\%$ for low-mass stars in this
cluster. By comparing our new mass-period relation for the Pleiades to
the slightly older cluster M35, we confirm previous indications that
the spin-down stalls at $\sim 100~{\rm Myr}$ for the slowest rotating
stars with $0.7~M_{\odot} \la M \la 1.1~M_{\odot}$ -- a fact which may
indicate that the internal transport of angular momentum is
inefficient in slowly rotating solar mass stars.
\end{abstract}


\section{Introduction}\label{sec:intro}

\begin{figure*}
\includegraphics[width=168mm]{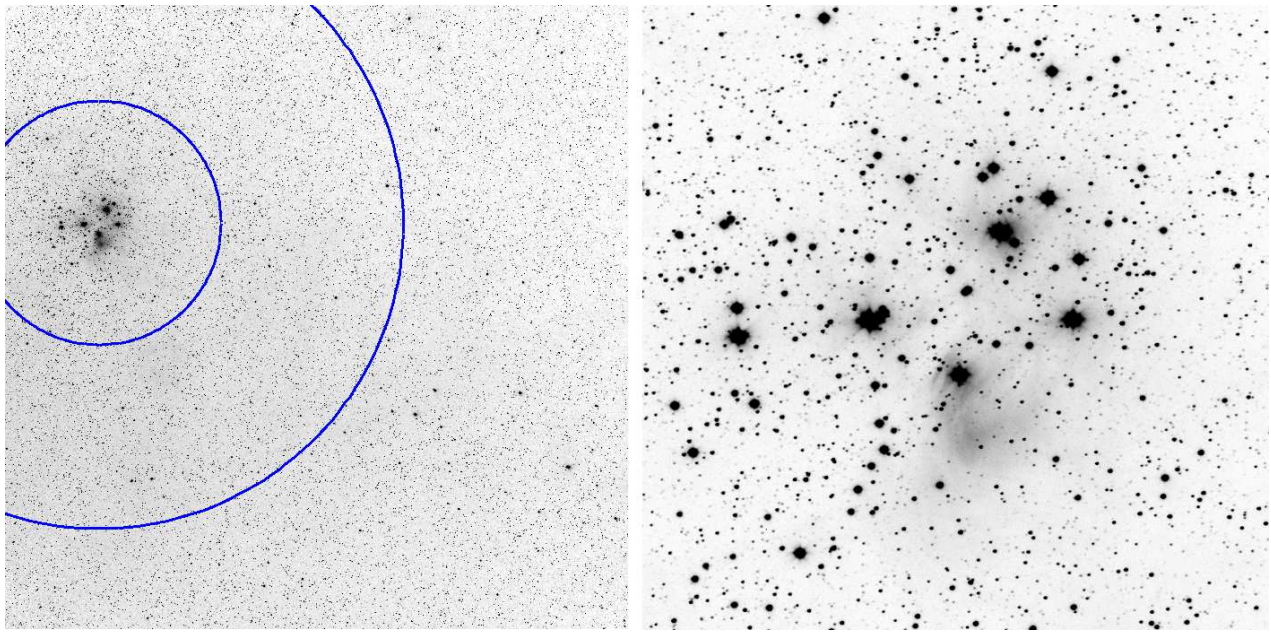}
\caption{ 
  Left: A typical $10.6\degr \times 10.6\degr$ HAT-9 image of
  HATNet field G259 containing the Pleiades cluster. The $2\degr$ and
  $5\degr$ radius circles contain 68\% and 100\% of the candidate
  cluster members in the catalogue of \citet{Stauffer.07}. Right: A
  $1\farcm5 \times 1\farcm5$ zoomed-in view of the center of the
  cluster.  }
\label{fig:fov}
\end{figure*}

Due to its proximity, richness, and age, the Pleiades star cluster ($d
\sim 133~{\rm pc}$, e.g.~\citealp{Soderblom.05}; ${\rm age} \sim
125~{\rm Myr}$, e.g.~\citealp{Stauffer.98}) has served for many years
as one of the benchmark clusters in studies of stellar angular
momentum evolution. \citet{Skumanich.72} first determined the
empirical spin-down relation for solar-type stars ($\omega \propto
t^{-1/2}$) by comparing measurements of the projected rotation
velocities ($v \sin i$) of solar-mass stars in the Pleiades and the
Hyades to the equatorial rotation velocity of the Sun. Since then
there have been numerous studies presenting rotation periods and $v
\sin i$ values for stars in the Pleiades (see the references given in
sections~\ref{sec:otherpercomp} and~\ref{sec:vsinicomp}).

Recently, large samples of stellar photometric rotation periods have
been published for a number of open clusters (see for example the
summary and references given by \citealp{IrwinBouvier.09} for 18 open
clusters, together with some new results, not included in their list,
for Coma Berenices by \citealp{CollierCameron.09}, for NGC~2301 by
\citealp{Sukhbold.09}, and for M34 by \citealp{James.10}). These data
have enabled theoretical investigations of the angular momentum
evolution of low-mass stars (again see the review by
\citealp{IrwinBouvier.09}, and also a recent study by
\citealp{Denissenkov.09}), which have led to a number of insights into
stellar physics. Observations of very young clusters with ages $\la
10~{\rm Myr}$ can be used to study phenomena such as accretion-driven
stellar winds \citep{Matt.05,Matt.08a,Matt.08b} or star-disk-locking
\citep[e.g.][]{Konigl.91,Shu.94}. Observations of clusters with ages
$\sim 50-200~{\rm Myr}$ are essential for constraining the time-scale
of coupling between stellar convective and radiative zones
\citep[e.g.][]{Bouvier.08,Denissenkov.09}. Finally, observations of
older clusters are important for constraining models of magnetized
stellar winds \citep{Kawaler.88} and for calibrating the rotation-age
relation \citep[e.g.][]{Barnes.07,Mamajek.08,CollierCameron.09}.

Although the comparison of models to the observed rotation period
distributions in open clusters has led to a number of insights, these
studies are hindered by incompleteness and poorly determined selection
effects in the available data-sets. These effects may in turn lead to
incorrect theoretical inferences. The available photometric rotation
periods for the Pleiades, in particular, show evidence of being biased
toward short periods \citep{Denissenkov.09}, this is unsurprising
since longer period stars generally have lower amplitude variations
and also require observations spanning a longer time-base to be
detected, moreover apparent variations due to systematic errors in the
photometry may dominate on long time-scales; the results to be
discussed here do not suffer from this bias because of the long
time-base and high precision of our observations. It is only possible
to determine whether or not the photometric periods are biased for this
cluster because it is fairly unique in having a rich, unbiased sample
of $v \sin i$ values in the literature. Given the importance of the
Pleiades for the study of stellar angular momentum evolution, an
unbiased sample of rotation periods for stars in this cluster would be
quite useful.

In this paper we present photometric periods that we associate with
rotation periods for \NUMperpleiadsfoundbyhat{} Pleiades stars. This
increases the number of Pleiades stars with measured periods by a
factor of 5, and importantly, it is 93\% complete for stars in the
mass range $0.7~M_{\odot} \la M \la 1.0~M_{\odot}$. For this survey we
use data from the Hungarian-made Automated Telescope Network
\citep[HATNet][]{Bakos.04}, a project that uses a network of 6
small-aperture, wide-field, fully-automated telescopes to search for
transiting exoplanets orbiting bright stars
\citep[e.g.][]{Bakos.10}. A similar study of stellar rotation for
field K and M dwarf stars using HATNet has been recently presented by
our group \citep{Hartman.09b}.

The structure of the paper is as follows: in section~\ref{sec:obs} we
briefly describe the photometric observations and data reduction; in
section~\ref{sec:persearch} we select periodic variable stars,
estimate the errors on the period, correct the light curves for
distortions, present the catalogue of rotation periods, and select new
cluster members; in section~\ref{sec:discussion} we compare our data
to previous rotation period and $v \sin i$ measurements, and we also
compare our results to other open clusters. We summarize the results
in section~\ref{sec:summary}.

\section{Observational Data and Initial Reduction}\label{sec:obs}

\begin{figure}
\includegraphics[width=84mm]{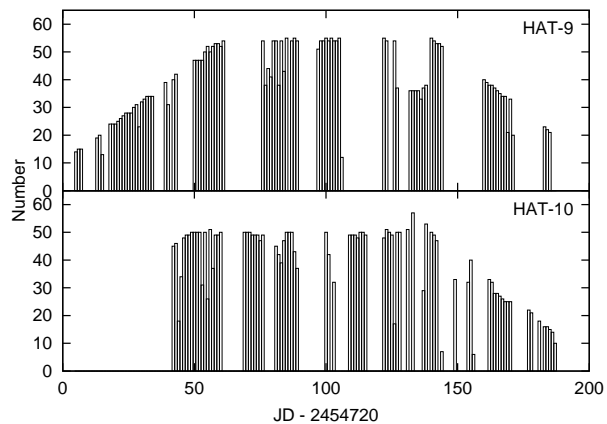}
\caption{ The number of observations of HATNet field G259 obtained on
  each night with the HAT-9 and HAT-10 instruments.  }
\label{fig:obshist}
\end{figure}

\begin{figure}
\includegraphics[width=84mm]{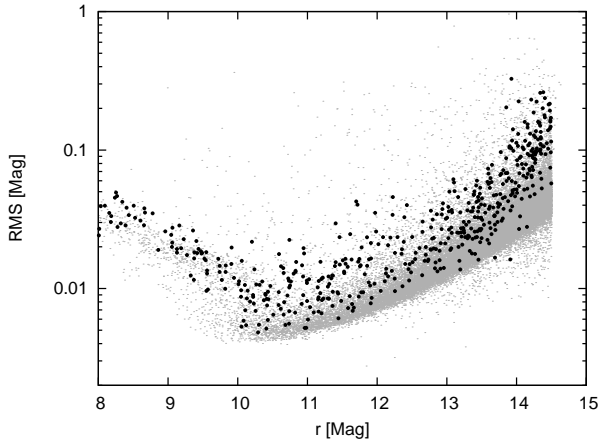}
\caption{ Unbiased RMS (eq.~\ref{eq:unbiasedrms}) vs. Sloan
  $r$ magnitude for the EPD/TFA corrected light curves of stars in
  field G259 (gray-scale points). The dark filled circles show
  probable Pleiades members (section~\ref{sec:catalogue}). Stars
  brighter than $r \la 10$ are saturated in a significant fraction of
  the images.}
\label{fig:lcstat}
\end{figure}

Photometric time-series observations of the Pleiades cluster were
obtained between 15 September 2008 and 16 March 2009 using the
identical HAT-9 and HAT-10 11 cm aperture robotic telescopes located
at Mauna Kea Observatory (MKO) in Hawaii and at Fred L. Whipple
Observatory (FLWO) in Arizona respectively. Each telescope used a
4K$\times$4K CCD and a Sloan $r$ filter to observe a $10.6\degr \times
10.6\degr$ field of view (FOV) centered at 03:30:00, 22:30:00
(J2000). The Pleiades are not centered in this field; the field
(internally designated as G259) was observed as part of the standard
operations of the HATNet transit survey, and was not specifically
chosen to observe the cluster. Figure~\ref{fig:fov} shows a typical
image of this field. The images have a pixel scale of $9\arcsec$; the
full-width at half-maximum (FWHM) of the point-spread function (PSF)
is $\sim 2.6$ pixels ($23\arcsec$). A total of 3648 exposures of 5
minutes, taken at 5.5 minute cadence, were obtained with the HAT-9
instrument, while 3138 images with the same exposure time and cadence
were obtained with the HAT-10 instrument. Figure~\ref{fig:obshist}
shows the temporal distribution of observations obtained with each of
these instruments.

The images were calibrated and reduced to light curves using tools
developed for the HATNet transit survey \citep[see][]{Pal.09,
  Bakos.04}. Briefly, after applying bias, dark current, and twilight
sky-flat calibrations in a standard fashion, stars were identified on
the images. The star lists were then matched to the Two Micron All-Sky
Survey \citep[2MASS;][]{Skrutskie.06} point source catalogue (PSC)
using the methods described by \citet{Pal.06} to determine the
astrometric solutions for the images. Aperture photometry is then
performed at the positions of all 2MASS sources with $r \la 14.5$
transformed to the image coordinate system. For each resulting light
curve the median magnitude is fixed to the approximate $r$ magnitude
of the source based on the following transformation from the 2MASS
$J$, $H$ and $K_{S}$ magnitudes \citep[see][ who give transformations
  from 2MASS magnitudes to Sloan $(g-r)$ and $(r-i)$ colors; we
  followed the same procedure to determine a transformation from 2MASS
  magnitudes to Sloan $r$ magnitude]{Bilir.08}:
\begin{equation}
r = 0.6975 + 2.9782J - 0.8809H - 1.1230K_{S}.
\end{equation}
The RMS scatter between the observed and predicted $r$ magnitudes for
the stars used to determine this relation is $0.076~{\rm mag}$. For
each source, photometry is performed using three apertures of radii
1.45, 1.95 and 2.35 pixels. Following the post-processing routines
discussed below, we adopt a single ``best'' aperture for each light
curve. For stars with $r > 13.5$ we take the smallest aperture to
minimize sky noise, for brighter stars we adopt the aperture for which
the processed light curve has the smallest RMS.

The initial ensemble calibration of the light curves against
variations in the flux scale was performed using the method described
in section 2.7.3 of \citep{Pal.09}. These light curves are then passed
through two routines that filter out systematic variations that are
correlated with measureable instrumental parameters or are present in
other stars in the field. The first routine, external parameter
decorrelation (EPD), decorrelates each light curve against a set of
measured instrumental parameters including the shape of the PSF, the
sub-pixel position of the star on the image, the zenith angle, the
hour angle, the local sky background, and the variance of the
background \citep[see][]{Bakos.10}. This decorrelation is done
independently on the data from the HAT-9 and HAT-10 telescopes. The
procedure is applied assuming that each star has a constant
magnitude. For large amplitude variable stars this may distort the
signal and may lower the S/N of the detection, however such large
amplitude variables will generally still be detectable.

After applying EPD, the light curves are then processed with the
Trend-Filtering Algorithm \citep[TFA;][]{Kovacs.05} which decorrelates
each light curve against a representative sample of other light curves
from the field. We used 530 template stars ($\sim 8\%$ of the total
number of images for the field). In applying the TFA routine we also
clip $5\sigma$ outliers from the light curves. At this point in the
analysis we apply EPD and TFA in signal-recovery mode (i.e. we apply
them under the assumption that the signal is constant), once a signal
is detected we then apply EPD and TFA in signal-reconstruction mode on
the original light curve to obtain an undistorted trend-filtered
light curve for the star (see section~\ref{sec:tfasr}). As for EPD,
signal-recovery mode TFA may distort the signal and lower the S/N of
large amplitude variable stars, though typically the effect is not
significant enough to prevent the variable from being selected.

In figure~\ref{fig:lcstat} we show the unbiased RMS of the EPD/TFA
corrected light curves for all stars in field G259. The unbiased RMS of a light curve is calculated using
\begin{equation}
{\rm RMS} = \sqrt{\frac{\sum \left( m_{i} - \langle m \rangle \right)^{2}}{N - N_{\rm P}}}
\label{eq:unbiasedrms}
\end{equation}
where $m_{i}$ are the individual magnitudes, $\langle m \rangle$ is
the average magnitude, $N$ is the number of points in the light curve,
and $N_{\rm P} = 544$ is the number of parameters used in applying
EPD/TFA. We mark separately the probable Pleiades members (see
section~\ref{sec:catalogue}). Stars with $r \la 10$ are saturated in a
significant fraction of the images (due to vignetting the exact
magnitude of saturation depends on the position of the star on the
image; it also depends on the sky brightness and transparency which
changes from image to image). The unbiased RMS of stars with $r \sim
10$ is $\sim 4~{\rm mmag}$. The light curves of Pleiades members
generally have greater RMS than those of field stars; this reflects
the fact that detectable photometric variability is significantly more
common for Pleiades members than it is for field stars (see
section~\ref{sec:catalogue}).

\section{Search for Periodic Variables}\label{sec:persearch}

We use the Lomb-scargle periodogram
\citep{Lomb.76,Scargle.82,Press.89} as implemented in the VARTOOLS
program \citep{Hartman.08b} to search the light curves for periodic
variations. We generate the periodogram for each light curve at a
frequency resolution of $0.1/T$ between $0.01~{\rm d}^{-1}$ and
$10.0~{\rm d}^{-1}$, where $T$ is the time-span of a given light
curve. The high-frequency cut-off of $10.0~{\rm d}^{-1}$ is adopted as
stars are not expected to have rotation periods shorter than this, and
expanding the range of the frequency search increases the bandwidth
penality, decreasing the significance of a given detection. We correct
the periodogram for red noise by removing any low-order frequency
dependence in the mean value of the periodogram and in the RMS of the
periodogram. To do this we fit a fifth order polynomial to the
periodogram, as well as a fifth order polynomial to the RMS of the
periodogram calculated in 100 frequency bins. We then define a new
periodogram using
\begin{equation}
p^{\prime}(\omega) = \frac{\langle \sigma \rangle}{\sigma(\omega)}(p(\omega) - \bar{p}(\omega)) + \langle \bar{p} \rangle
\label{eqn:rncorrect}
\end{equation}
where $p$ is the value of the original periodogram at frequency $\omega$,
$\sigma(\omega)$ is the polynomial fit to the RMS of the periodogram
as a function of frequency, $\bar{p}(\omega)$ is the polynomial fit to
the periodogram, and the brackets denote averaging over frequency. We
perform the fit using 5-$\sigma$ iterative clipping, using the model
value of $\sigma(\omega)$ in the clipping.

\begin{figure}
\includegraphics[width=84mm]{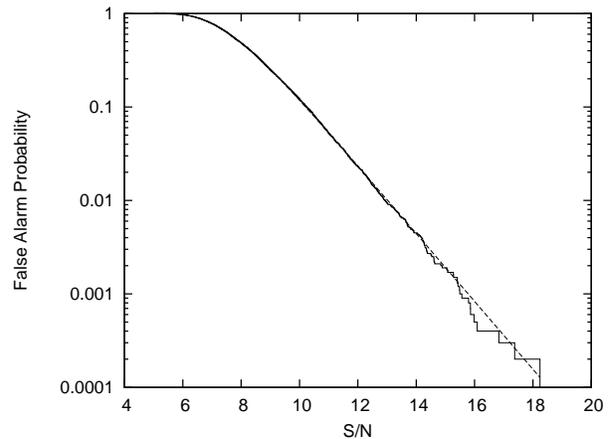}
\caption{False alarm probability vs. S/N of the peak in the red-noise corrected L-S periodogram. The solid line shows the results from conducting simulations of white noise light curves, the dashed lines shows eq.~\ref{eq:lssnfap}.}
\label{fig:LSSNFAP}
\end{figure}

We identify the highest peak in the corrected periodogram $p^{\prime}$
and determine its S/N using an iterative 5-$\sigma$ clipping in
calculating the RMS of the periodogram. To set the selection threshold
we simulate 10000 white noise light curves with the same time sampling
as the observations, calculate the red-noise corrected L-S periodogram
of each simulation, and determine the S/N of the highest peak in each
periodogram using the same procedure as used for the real light
curves. Figure~\ref{fig:LSSNFAP} shows the false alarm probability
(FAP) as a function of $S/N$ determined from the simulations, which we
find to be well-fit by a function of the form:
\begin{equation}
{\rm FAP} = 1 - (1 - e^{-0.84({\rm S/N} + 1)})^{1240}.
\label{eq:lssnfap}
\end{equation}
This is similar to the form expected for a normalized L-S periodogram
without red-noise correction or clipping \citep[e.g.][]{Press.92}. We
adopt a cut-off of S/N$> 16$ which corresponds to a 0.08\% FAP. A
total of 2236 of the 36,011 stars in the field pass the selection. In
section~\ref{sec:catalogue} below we focus on the stars that are
likely to be cluster members. Adopting a lower threshold of S/N$> 13$
(corresponding to a 1\% FAP) would increase the total number of
selected potential variable stars to 2878.

\subsection{Resolving Aliases}\label{sec:alias}

In some cases there is an ambiguity in choosing the correct period
from among several possible aliases or harmonics. In general we follow
the convention of adopting the highest S/N peak in the red-noise
corrected L-S periodogram. We found, however, that often the red-noise
correction changes the most signficant peak in the periodogram to a
high frequency alias $|f_{\rm s} - f_{0}|$ of the most significant
peak in the uncorrected periodogram, $f_{0}$. Here $f_{\rm s}$ is the
sidereal frequency. In these cases we adopt the period from the
uncorrected periodogram. We found that generally the periods from the
red-noised corrected periodograms pile-up near $1~{\rm day}$ while the
periods from the uncorrected periodograms fall along the main
period-mass relation (see figure~\ref{fig:permass}).

\subsection{Estimation of Period Uncertainties}\label{sec:pererr}

As discussed by \citet{Hartman.09a}, the primary factors that
contribute to errors in the measured photometric rotation period are:
\subsubsection*{Instrumental Effects}
\begin{enumerate}
\item choosing an alias or harmonic of the true period,\label{enum:aliaserr}
\item noise in the photometry together with finite sampling of the
  light curve,\label{enum:noiseerr}
\item inadequacies in the model used to determine the period (e.g.,
  the light curve is periodic but not sinusoidal),\label{enum:badmodelerr}
\end{enumerate}
\subsubsection*{Physical Effects}
\begin{enumerate}
\setcounter{enumi}{3}
\item spot evolution, and\label{enum:spotevolveerr}
\item differential rotation.\label{enum:diffroterr}
\end{enumerate}

The periodogram for a light curve will contain several discrete peaks
at harmonics and aliases of the true period. Deciding which peak is
the correct one is difficult to do in general. We do not consider this
effect in estimating the uncertainties on the measured periods. 

The effects \ref{enum:noiseerr}-\ref{enum:spotevolveerr} listed above
all contribute to the spread in the periodogram peak, which reduces
the precision with which the position of the peak may be measured. We
determine the resulting error in the period by fitting a Gaussian
function to the points within $2/T$ of the periodogram peak and
measuring the 1$\sigma$ spread in the position of the peak using the
Extended Markov-Chain Monte Carlo technique \citep[see][]{Pal.09}. We
set the error for each point in the periodogram equal to the RMS of
the full periodogram determined with an iterative $5\sigma$
clipping. We also record the standard deviation of the best-fit
Gaussian, as this contains information on the spot life-times.

Differential rotation may contribute to the uncertainty in two
ways. If there are multiple spot groups at different stellar
latitudes, the differing rotation periods for each group will broaden
the periodogram peak, and will contribute to the uncertainty in the
period as just described. Differential rotation will also lead to a
systematic error in determining the rotation period of the star since
the degree of differential rotation and the latitude of the dominant
spot group are not known, making it impossible to relate the measured
period to the period at a reference latitude (such as the equator). If
Pleiades stars exhibit solar-like differential rotation (the rotation
period increases towards the poles), the measured periods will be
systematically longer than the equatorial period. The Sun exhibits
differential rotation following
\begin{equation}
P_{\beta} = P_{\rm EQ}/(1 - k \sin^{2}\beta)
\label{eq:diffrot}
\end{equation}
where $P_{\beta}$ is the rotation period at latitude $\beta$, $P_{\rm
  EQ}$ is the equatorial rotation period, and $k = 0.19$ when using
spots to track the rotation \citep{Kitchatinov.05}. For younger, more
rapidly rotating stars, the value of $k$ is expected to decrease
\citep{Brown.04}. Observations of the $P = 8.77~{\rm d}$ solar-like
star $\kappa^{1}$~Ceti by the {\em Microvariability and Oscillations
  of Stars (MOST)} satellite bear this theoretical expectation out,
finding $k = 0.09$ \citep{Walker.07}. The observations of
$\kappa^{1}$~Ceti also indicate that spots on rapidly rotating stars
may be found at any latitude; this is in contrast to the Sun where
spots are rarely seen with $|\beta| > 30\degr$. Assuming the dominant
spot groups are isotropically distributed on Pleiades stars
(i.e. $\sin \beta$ is uniformly distributed), and that these stars
exhibit solar-like differential rotation with $k = 0.09$, we expect
the measured rotation periods to have a mean value of $1.03 P_{\rm
  EQ}$, with a standard deviation of $0.03 P_{\rm EQ}$. For $k = 0.19$
the respective values are $1.07 P_{\rm EQ}$ and $0.07 P_{\rm EQ}$. We
do not include systematic errors due to differential rotation in the
period uncertainties reported in our catalogue.

\subsection{Signal Reconstruction with EPD/TFA}\label{sec:tfasr}

Once a period is determined for a star, we obtain a new trend-filtered
light curve by running EPD and TFA in signal reconstruction mode
\citep[e.g.][]{Kovacs.05}. This correction is important to get an
unbiased measurement of the amplitude of photometric variations. To do
this, we fit to the pre-EPD light curve (i.e. the light curve has been
corrected for ensemble variations in the flux scale, but has not had
any other filtering applied to it), a model of the form:
\begin{eqnarray}
m(t) & = & a_{0} + \sum_{i=1}^{10}\left( a_{i}cos(2\pi i t / P) + b_{i}sin(2\pi i t / P)\right) \nonumber \\
& & + \sum_{i=1}^{N}c_{i} s_{i}(t).
\label{eqn:epdtfasr}
\end{eqnarray}
The first sum is a Fourier series with period $P$ (the period found
for the light curve in section~\ref{sec:persearch}) which is used to
represent the physical signal, while the second sum is a model for the
instrumental/atmospheric variations. The free parameters in this model
are the $a_{i}$, $b_{i}$ and $c_{i}$ coefficients, while the
$s_{i}(t)$ terms, which represent instrumental variations, consist of
530 template light curves, together with 14 known instrumental
parameter sequences (including: the X and Y subpixel positions of the
star to first and second order, 3 parameters describing the PSF shape
to first and second order, the hour angle, the zenith angle, the
background, and the deviation of the background). Because all free
parameters enter linearly in eq.~\ref{eqn:epdtfasr}, the fit can be
done quickly using singular value decomposition
\citep[e.g.][]{Press.92}. 

We take the amplitude of the light curve to be equal to the
peak-to-peak amplitude of the Fourier series in
eq.~\ref{eqn:epdtfasr}. For four stars, HAT-259-0000690 (PELS~020),
HAT-259-0000923 (HII~120), HAT-260-0000924 (HII~1182) and
HAT-260-0007928 (HII~906), with periods very close to $3$, $4$, $3$
and $2$ days respectively, the poor phase coverage of the HATNet light
curves means that the high order harmonics in the Fourier series are
not well constrained. For these stars we therefore do not include any
harmonics other than the fundamental (i.e. the Fourier series includes
one term only) when fitting equation~\ref{eqn:epdtfasr}.

The median value of the ratio of amplitudes from signal {\em
  reconstruction} EPD/TFA light curves to amplitudes from signal {\em
  recovery} EPD/TFA light curves is 1.3. This illustrates the
importance of using signal reconstruction EPD/TFA to get an unbiased
amplitude measurement.

\begin{figure}
\includegraphics[width=84mm]{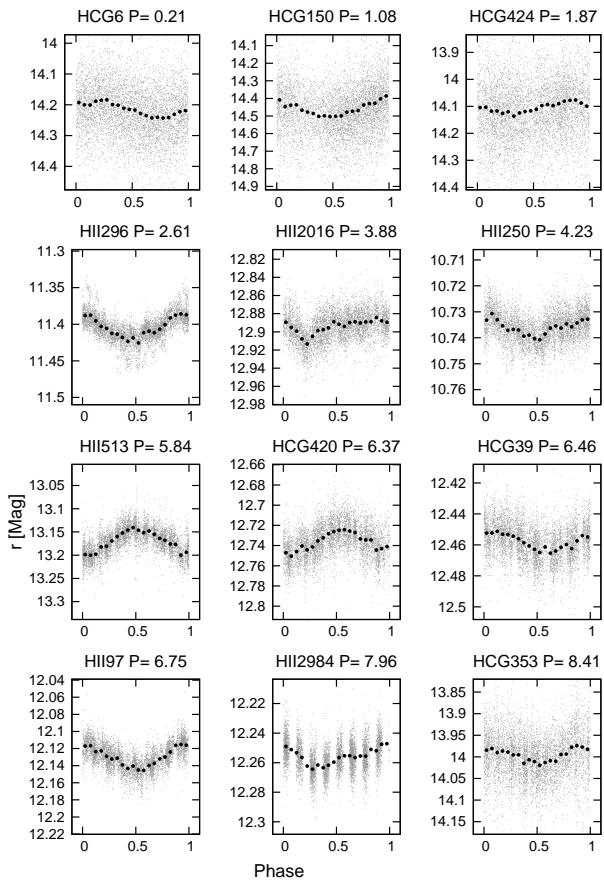}
\caption{Phased light curves for 12 stars randomly selected from our
  catalogue, sorted by rotation period. The gray-scale points show all
  the photometric data, the dark filled circles show the phase-binned
  light curve. The name for each star is taken from
  \citet{Stauffer.07} and the listed period is in days. These light
  curves have had signal-reconstruction EPD/TFA applied
  (Section~\ref{sec:tfasr}).}
\label{fig:examplelcs}
\end{figure}

\begin{figure}
\includegraphics[width=84mm]{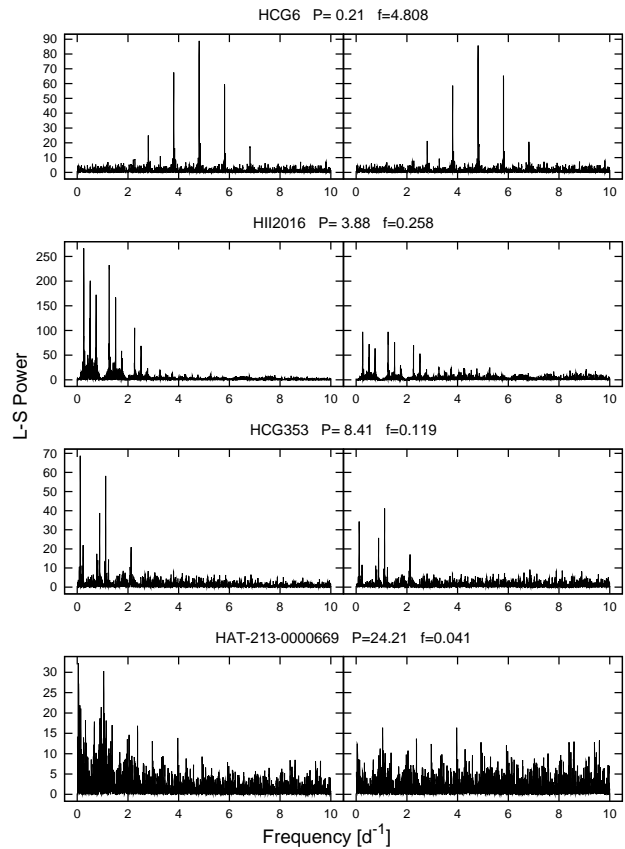}
\caption{L-S periodograms for 3 of the stars displayed in
  figure~\ref{fig:examplelcs}, and for a fourth field star which
  passes the S/N selection before applying the red-noise correction to
  the periodogram (equation~\ref{eqn:rncorrect}), but does not pass
  the selection after applying the correction. For each star we show
  the original L-S periodogram on the left, and the red-noise
  corrected periodogram on the right. We list the period (in days) and
  the frequency (in cycles per day) selected for each star. For the
  star in the bottom panel, this is the period that is selected from
  the un-corrected periodogram. For HCG~353 the low frequency alias is
  adopted following the procedure described in
  Section~\ref{sec:alias}.}
\label{fig:examplespec}
\end{figure}

\subsection{Catalogue of Rotation Periods for Pleiades Members}\label{sec:catalogue}

Using the recent compilation of probable Pleiades members by
\citet[][hereafter S07]{Stauffer.07}, we identify \NUMpleiads{}
probable members with magnitudes in the range $9.5 < r < 14.5$ for
which we have obtained light curves. We detect periodic variations for
\NUMperpleiads{} of these stars (i.e. 74\% of them); for stars with
$11 < r < 13$, our period detection rate is 93\%. Faint stars with $13
< r < 14.5$ have low photometric precision making it more difficult to
detect their variations, while hotter stars with $9.5 < r < 11$ have
lower amplitudes of variation. Stars with $9.5 < r < 10$ may also be
saturated in a significant fraction of the images. As a comparison,
for stars with $11 < r < 13$ that are not selected as probable cluster
members, the period detection rate is 9.8\%. Two factors that lead to
a much higher incidence of variability for Pleiades stars include
their young age relative to field stars (so that they are more active
and have higher amplitudes of variability), and the fact that the
majority of field stars are more distant than the Pleiades, so
Pleiades stars of a given mass are brighter, and have higher precision
light curves, than most field stars of that mass in the survey.

Our final catalogue of rotation periods is included in the
supplementary material to the online edition of this
article. Table~\ref{tab:prdcat} shows a portion of the catalogue for
guidance regarding its form and content. We include in the catalogue
fifteen stars with rotation periods in the literature that we do not
recover, including the star CFHT-PL~8 which was not included in the
S07 catalogue, and exclude one star with a period in the literature
that is not a cluster member (see Section~\ref{sec:otherpercomp} for
further details on these stars). We also include
\NUMpleiadsnotinstauffer{} variable stars which are not in the S07
catalogue, but which have photometry, proper motions, and periods
consistent with cluster membership (see section~\ref{sec:newmembers}
below).

Figure~\ref{fig:examplelcs} shows phased light curves for 12 stars
randomly selected from our catalogue, while
figure~\ref{fig:examplespec} shows L-S periodograms for 3 of these
stars. Figure~\ref{fig:examplespec} also shows an example of a star
which is not selected after applying the red-noise correction to the
periodogram (equation~\ref{eqn:rncorrect}). The photometric light
curves for all \NUMpleiadstot{} probable members observed with HATNet
will be made publicly available from the NASA Star and Exoplanet
Database
\citep[NStED;][]{vonBraun.09}\footnote{http://nsted.ipac.caltech.edu}.

In addition to the catalogue of rotation periods, we also provide a
catalogue of Pleiades stars observed with HATNet for which we do not
detect a period (Table~\ref{tab:memnonvarcat}), a catalogue of field
variable stars (Table~\ref{tab:fieldprd}), and a list of previous
period measurments for Pleiades stars (Table~\ref{tab:knownprd}).

\subsection{New Cluster Members}\label{sec:newmembers}

Here we leverage the enhanced photometric variability of Pleiades
stars relative to field stars to identify new members of the
cluster. A similar method for selecting Pleiades members was employed
by \citet{Haro.82} who identified flare stars as candidate
members. More recently, \citet{CollierCameron.09} used photometric
variability to select members of the Coma Berenices open cluster.

\begin{figure}
\includegraphics[width=75mm]{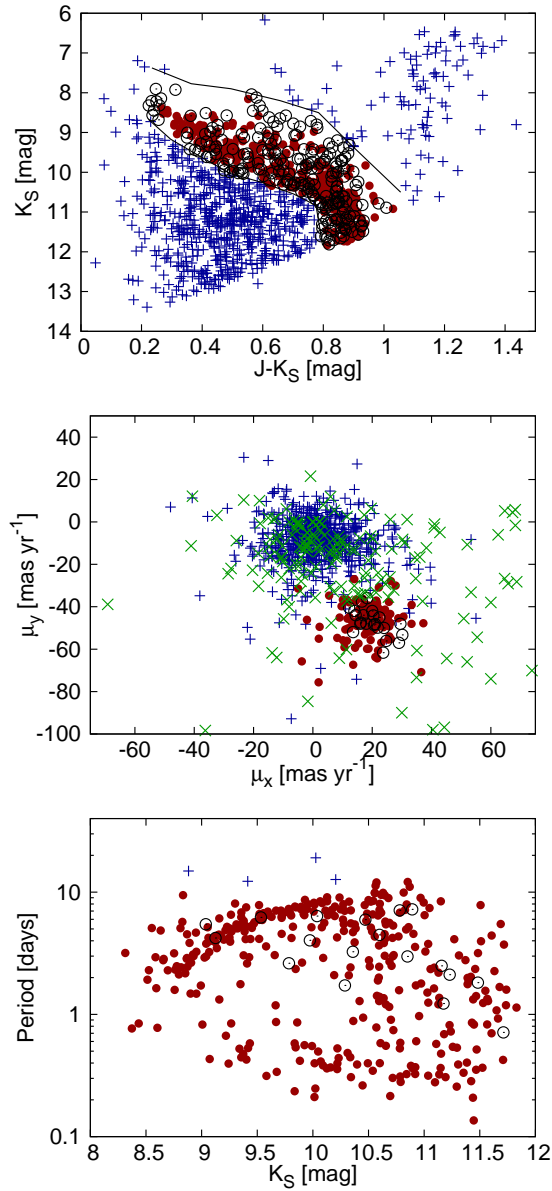}
\caption{ Top: $J-K_{S}$ CMD for all variable stars. Cluster members
  from S07 are plotted with filled circles,
  variables with $J-K_{S}$ photometry consistent with cluster
  membership are plotted with open circles, and variables that
  are not photometric members are plotted with crosses. The solid
  lines dilineate the selection of candidate cluster
  members. Centre: Vector Point Diagram (VPD) for all identified
  variable stars. Filled circles show cluster members from
  S07, crosses are stars which were rejected as
  non-photometric members in the left panel, open circles show
  photometric members which also have a proper motion membership
  probability $> 70\%$, Xs show photometric members which have a
  proper motion membership probability $< 70\%$. Bottom: Period
  vs. $K_{S}$-magnitude. Filled circles show cluster members from
  S07, open circles correspond to new candidate
  members which passed the other two selections and also have periods
  that are consistent with cluster membership. Crosses show
  candidate members that passed the previous selections, but have
  periods that are too long to be cluster members. }
\label{fig:memselect}
\end{figure}

Figure~\ref{fig:memselect} shows the selection of cluster members. We
first select variable stars that are within $5\degr$ of the centre of
the cluster, and are near the cluster main sequence on a $J-K_{S}$
colour-magnitude diagram (CMD), taking the photometry from the 2MASS
$6\times$ PSC (see the description by Cutri et
al.\footnote{http://www.ipac.caltech.edu/2mass/releases/allsky/doc/seca3\textunderscore1.html})
when available, and from the main 2MASS PSC for all other stars. A
total of 221 of the 958 variables not included in the S07 catalogue
that are within $5\degr$ of the cluster centre pass this selection. Of
these, 189 have proper motions from the PPM-Extended catalogue
\citep{Roser.08}, while 28 of the remaining 32 stars have proper
motions from the USNO-B1.0 catalogue \citep{Monet.03}. To establish
the proper motion membership probability we follow the procedure
outlined in \citet{Deacon.04}. Briefly, we assume that proper motions
are distributed as:
\begin{equation}
\Phi = f\Phi_{f} + (1-f)\Phi_{c}
\end{equation}
where $f$ is the fraction of stars in the field, $\Phi_{f}$ is the
distribution for field stars, and $\Phi_{c}$ is the distribution for
cluster members. We assume bivariate Cauchy distributions for the
field and the cluster of the form:
\begin{eqnarray}
\Phi_{f} & = & \frac{1}{2\pi\Gamma_{f}^{2}}\frac{1}{\left(1+((\mu_{x}-\mu_{xf})^{2}+(\mu_{y}-\mu_{yf})^2)/\Gamma_{f}^2\right)} \\
\Phi_{c} & = & \frac{1}{2\pi\Gamma_{c}^{2}}\frac{1}{\left(1+((\mu_{x}-\mu_{xc})^{2}+(\mu_{y}-\mu_{yc})^2)/\Gamma_{c}^2\right)}
\end{eqnarray}
where $\mu_{x}$ and $\mu_{y}$ are the proper motion in the right
ascenscion and declination directions respectively. We determine the
free parameters $f$, $\mu_{xf}$, $\mu_{yf}$, $\mu_{xc}$, $\mu_{yc}$,
$\Gamma_{f}$ and $\Gamma_{c}$ by using the downhill simplex algorithm
to maximize the likelihood function
\begin{equation}
L = \sum_{i} \ln \Phi_{i}
\end{equation}
where the sum is over all variable stars. We also attempted to use
Gaussian and Exponential distributions for $\Phi_{f}$ and $\Phi_{c}$,
but found that the Cauchy distributions provided the best match to the
observations. The membership probability for a star is then given by
\begin{equation}
p = \frac{(1-f)\Phi_{c}}{f\Phi_{f} + (1-f)\Phi_{c}}.
\end{equation}
We require $p > 70\%$ for the star to be considered a probable member.
Of the 22 stars which pass the above selections, we reject 4 with
periods longer than any known cluster members of comparable magnitude.

Four of the \NUMpleiadsnotinstauffer{} remaining probable cluster
members have previously been selected as potential cluster members,
but were not included in the S07 catalogue. These include HCG~84 and
HCG~235, two well-known flare stars included in the catalogue of
Pleiades members by \citet{Haro.82}, and in several subsequent
membership catalogues, and the stars SRS~79807 and SRS~34337 which
were identified as proper-motion members by \citet{Schilbach.95}, but
to our knowledge, have not appeared in subsequent studies. All
\NUMpleiadsnotinstauffer{} stars are included in the rotation period
catalogue.

\section{Discussion}\label{sec:discussion}

\subsection{Comparison With Other Period Measurements}\label{sec:otherpercomp}

To compare our measurements to previous observations of the Pleiades,
we use the compilation of rotation periods provided on the WEBDA
database\footnote{http://www.univie.ac.at/webda/webda.html}. The
original measurements come from a variety of sources
\citep{Stauffer.87, Prosser.93a, Prosser.93b, Prosser.95, Marilli.97,
  Krishnamurthi.98, Terndrup.99, Messina.01a, Clarke.04,
  Scholz.04}. We also include periods for 11 stars from
\citet{VanLeeuwen.87} that are not listed on WEBDA. A total of
\NUMperprevknown{} stars are included in the compilation, which is
provided in table~\ref{tab:knownprd}.

We recover rotation periods for \NUMperprevknownperfoundmatch{} of the
stars with previous measurements. Here we discuss the 16 stars with
previous measurements which we do not recover. The
\NUMperprevknowntoofaint{} very low-mass stars studied by
\citet{Scholz.04} are all too faint to be observed with HATNet, the
star HD~23386 studied by \citet{Marilli.97} and \citet{Messina.01a} is
saturated in our images. There are four stars that we observed but did
not detect a period for, including: HII~708
\citep{Prosser.93b,Marilli.97}, HII~727
\citep{Prosser.93b,Marilli.97}, HII~975 \citep{Marilli.97} and HHJ~409
\citep{Terndrup.99}. The first three of these stars are near the
bright end of the magnitude range covered by HATNet, while HHJ~409 is
near the faint end. Finally there are two stars listed as variables on
WEBDA that are not included in the S07 catalogue. One of these stars,
HII~3167, was reported to be variable by \citet{Clarke.04} with a
$0.25~{\rm mag}$ amplitude in $B$, $0.15~{\rm mag}$ amplitude in $V$
and a period of $\sim 0.9~{\rm d}$. We find no evidence for
variability in our light curve for this star \citep[we match the star
  to 2MASS~J2000.0351555+2442326 based on the plate x/y position given
  on WEBDA, however this identification may not correspond to the same
  star observed by][]{Clarke.04}. In any case the star is not a
cluster member \citep[based on the color and magnitude given
  by][]{Clarke.04}, so we do not include it in the final catalogue of
Pleiades variables. The other known variable that is not in the S07
catalogue is CFHT-PL~8 \citep{Bouvier.98}, which was identified as a
variable by \citet{Terndrup.99}. This very low mass star is too faint
to be observed by HATNet. We include it in the final catalogue of
rotation periods, with coordinates and $K_{S}$ photometry taken from
the 2MASS catalogue, and $V$ magnitude taken from \citet{Samus.03}.

\begin{figure}
\includegraphics[width=84mm]{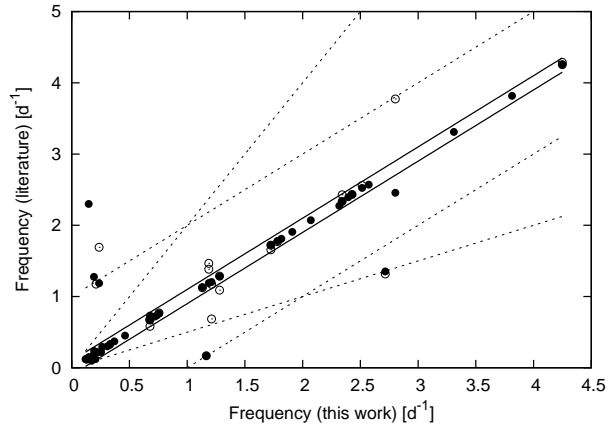}
\caption{Comparison between the rotation frequencies measured in our
  survey and those given in the literature for stars with previous
  measurements. For stars with multiple frequencies given in the
  literature, we show the literature value that is closest to our own
  (filled circles); the open circles show the other literature values
  for these stars. Points between the solid lines have literature
  frequencies that are within 10\% of our frequencies. The dotted
  lines show the $\pm 1~{\rm d}^{-1}$ alias frequencies, and the $2.0$
  and $0.5$ harmonic frequencies.}
\label{fig:FreqCompare}
\end{figure}

For \NUMperprevknownperfoundmatchclosefreq{} of the
\NUMperprevknownperfoundmatch{} previously studied stars that we
recover, our frequency is within $0.1~{\rm d}^{-1}$ of at least one of
the previously published frequencies for the star; and for
\NUMperprevknownperfoundmatchrealclosefreq{} stars our frequency is
within $0.01~{\rm d}^{-1}$ of at least one of the previously published
frequencies. Figure~\ref{fig:FreqCompare} shows a comparison of the
rotation frequencies measured in our survey to those given in the
literature for stars with previous measurements. Below we discuss the
6 cases where our frequencies are substantially different from all of
the literature values:

\emph{ HII~250} - Has published periods of $0.843~{\rm days}$ from
\citet{Marilli.97} and $0.591~{\rm days}$ from \citet{Messina.01a}. We
measure a period of $\sim 4.232~{\rm days}$. Our light curve shows no
evidence for periodicity at either of the published periods.

\emph{HII~335} - Has published periods of $0.265~{\rm days}$ from
\citet{Stauffer.87} and $0.4073~{\rm days}$ from
\citet{Messina.01a}. \citet{Stauffer.87} also report an alias period
of $0.360~{\rm days}$ for this star that is consistent with our
measured period of $0.357~{\rm days}$.

\emph{HII~885} - The only published period for this star is
$0.435~{\rm days}$ from \citet{Marilli.97}. We measure a period of
$6.83~{\rm days}$, and find no evidence for periodicity at $0.435~{\rm
  days}$.

\emph{HII~1039} - The only published period for this star is
$0.784~{\rm days}$ from \citet{Messina.01a}. We measure a period of
$5.22~{\rm days}$, and find no evidence for periodicity at $0.784~{\rm
  days}$.

\emph{HII~1124} - Has published periods of $5.9~{\rm days}$ from
\citet{Prosser.95} and $6~{\rm days}$ from \citet{VanLeeuwen.87}. We
find a period of $0.858~{\rm days}$ for this star, however prior to
correcting the periodogram for red-noise the top peak in the
periodogram for this star is at $6.133~{\rm days}$. This star has $v
\sin i=3.50$\,km\,s$^{-1}$ \citep{Queloz.98}, which is closer to the
expected equatorial rotation velocity of $\sim 5.8$\,km\,s$^{-1}$ for
the $\sim 6~{\rm day}$ period than the velocity of $\sim
40$\,km\,s$^{-1}$ for the shorter period. We therefore adopt the
longer period for the final catalogue.

\emph{HII~1653} - Both \citet{Krishnamurthi.98} and
\citet{Messina.01a} measured periods of $\sim 0.75~{\rm days}$ for
this star. Our period of $0.368~{\rm days}$ is half the value of the
published periods, and is inconsistent with the previously published
light curves for the star. We therefore double our period for the
final catalogue.

\subsection{Comparison with $v \sin i$ data}\label{sec:vsinicomp}

\begin{figure}
\includegraphics[width=84mm]{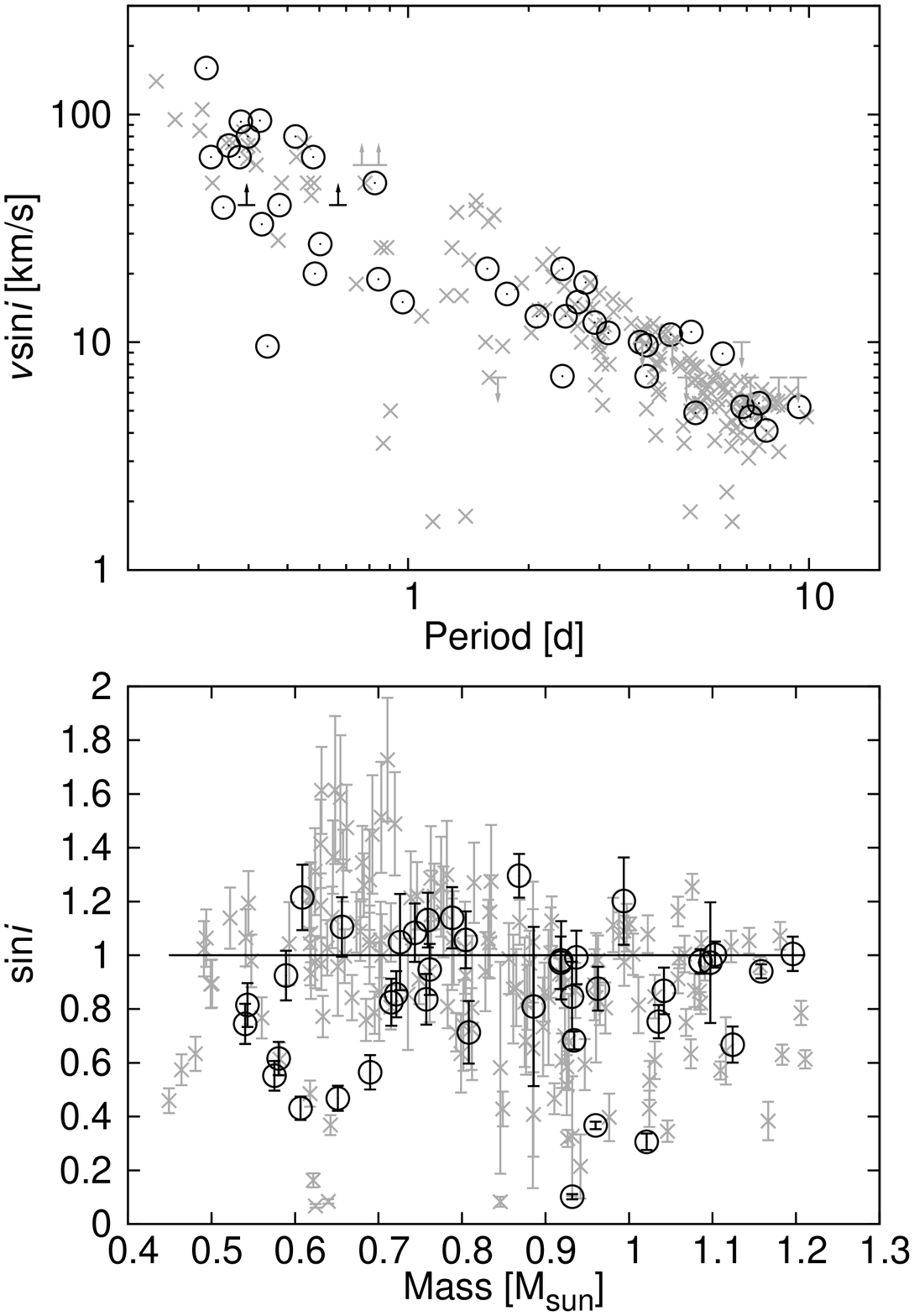}
\caption{Top: The spectroscopically determined $v \sin i$ taken from
  the literature vs. the photometric rotation period determined in
  this paper for \NUMwithperandspec{} stars. Open circles show objects
  selected as photometric binaries and grey-scale Xs show
  non-binaries. Arrows show stars for which only upper or lower limits
  on $v \sin i$ are available. Bottom: $\sin i$ calculated from $v
  \sin i$, the rotation period, and the radii inferred from an
  isochrone, is plotted against stellar mass inferred from an
  isochrone (see section~\ref{sec:vsinicomp} for details). Stars
  falling above the line $\sin i = 1$ have measured rotation period,
  $v \sin i$ and stellar radius values that are not fully consistent
  with one another. The error-bars include uncertainties in $v \sin
  i$, the period, and the radius (propagated from the uncertainty on
  $K_{s}$). We assume a 10\% uncertainty on $v \sin i$ for stars
  without $v \sin i$ uncertainties given in the literature. We exclude
  known and suspected spectroscopic binary systems from both plots.}
\label{fig:pervsini}
\end{figure}

\begin{figure*}
\includegraphics[width=168mm]{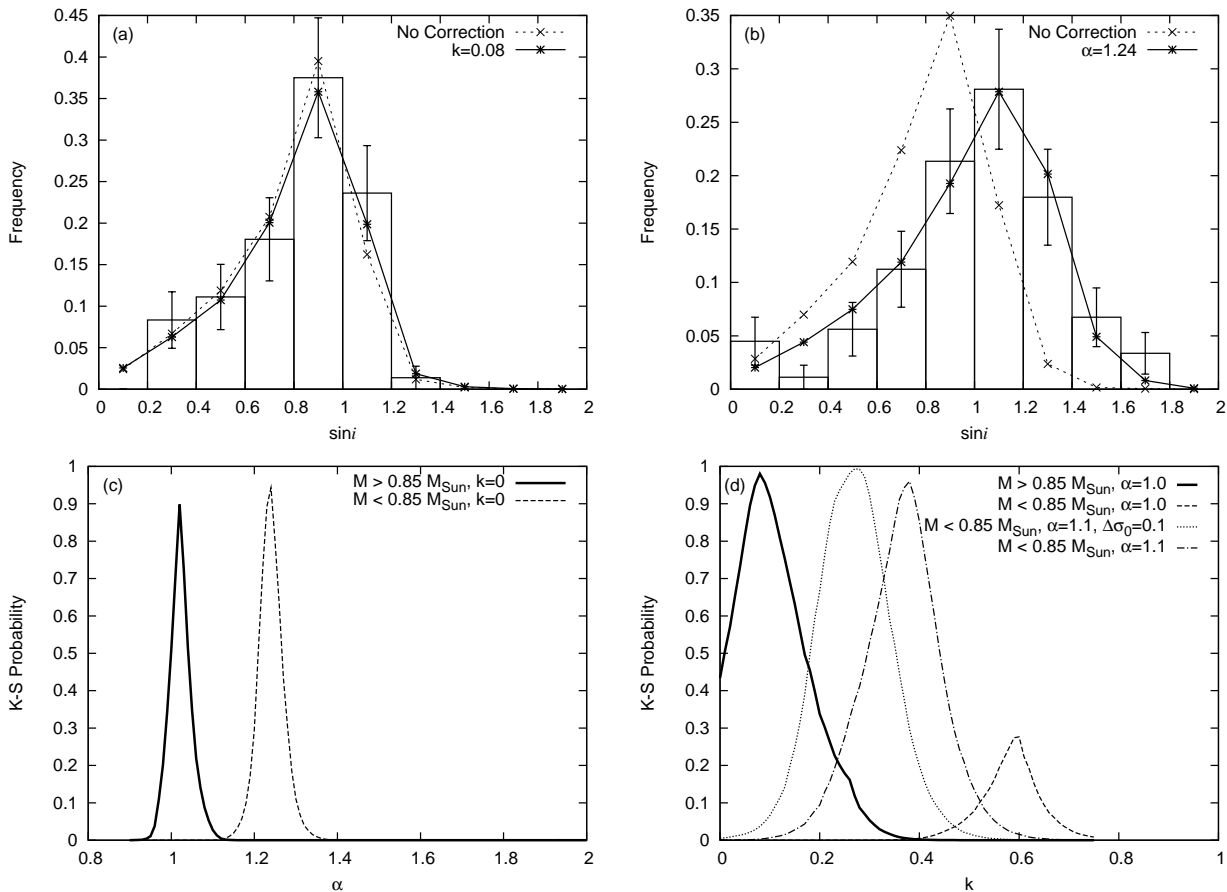}
\caption{The distribution of $\sin i$ values for stars with $M >
  0.85~M_{\odot}$ (a) and for stars with $M < 0.85~M_{\odot}$
  (b). These are compared to the expected distribution assuming stars
  have random rotation axis orientations and that there are no
  systematic biases in the measured parameters (the ``No Correction''
  line), as well as to a model that includes differential rotation
  (the
  $k=0.08$ line in panel (a)), and a model that includes a scale
  factor $\alpha$ applied to the $\sin i$ values (the $\alpha=1.24$ line
  in panel (b)). We also show the probability that the observed $\sin
  i$ distribution is drawn from a model distribution for models
  invoking a scale factor $\alpha$ (c) and for models that include
  differential rotation (d).}
\label{fig:sinisim}
\end{figure*}

\begin{figure}
\includegraphics[width=84mm]{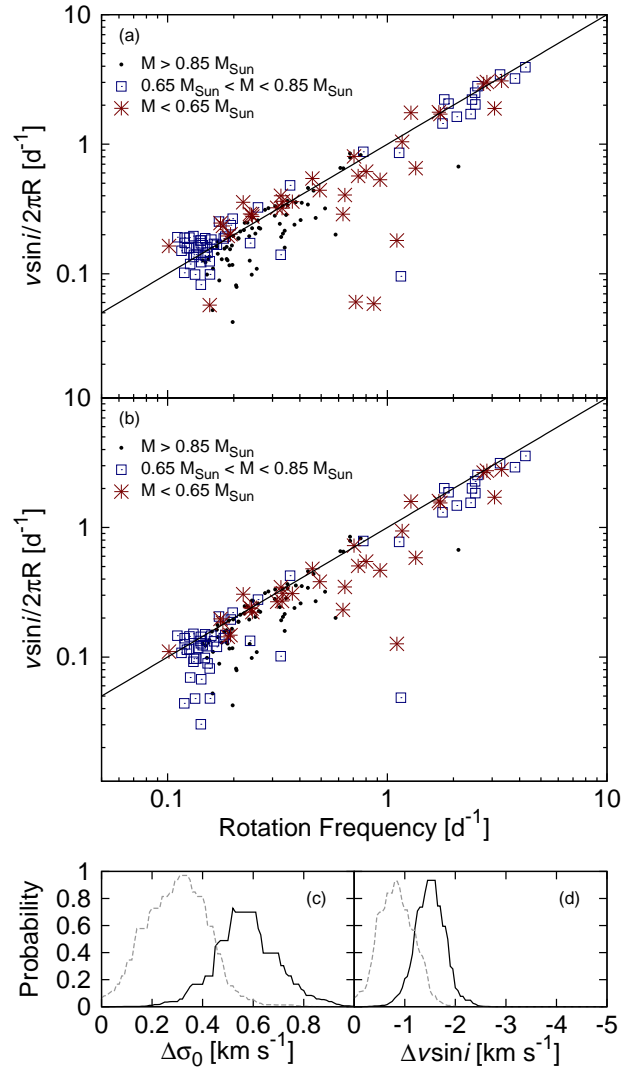}
\caption{a: Photometric rotation frequency ($1/P$) vs. the frequency
  inferred from $v \sin i$. The solid diagonal line shows the upper
  limit ($\sin i = 1$). The stars are divided into three mass
  bins. Low frequency (long period) stars with $M < 0.85~M_{\odot}$
  are more likely to lie above the upper limit than higher frequency
  stars with comparable masses. Stars with $M > 0.85~M_{\odot}$
  generally fall below the upper limit for all frequencies. b: Same as
  above, here the velocities of stars with $M < 0.85~M_{\odot}$ have
  been corrected with $\Delta \sigma_{0} = 0.16$\,km\,s$^{-1}$, and
  the radii of these stars have been increased by $10\%$, see the text
  for details. c: The probability that stars with $M < 0.85~M_{\odot}$
  have the same $\sin i$ distribution as stars with $M >
  0.85~M_{\odot}$ as a function of the applied $\Delta \sigma_{0}$
  correction. The solid line is for the case when no radius correction
  is applied, the dashed line shows the case when a 10\% correction is
  applied to the radii of stars with $M < 0.85~M_{\odot}$. d: Same as
  C, here a $\Delta v \sin i$ correction is applied.}
\label{fig:freqpervfreqvsini}
\end{figure}

We compare our period measurements to the rich sample of $v \sin i$
measurements available for members of the Pleiades. Measurements of $v
\sin i$ are taken from \citet{Stauffer.87}, \citet{Soderblom.93},
\citet{Queloz.98}, and \citet{Terndrup.00}. A total of
\NUMwithperandspec{} of the stars for which we detect periods also have
a $v \sin i$ value given in one of these sources. 

Figure~\ref{fig:pervsini} compares the photometric rotation periods
and the $v \sin i$ values, and also compares the inferred $\sin i$ to
the stellar mass. The results are shown separately for objects
selected as lying on the single-star main sequence, and objects
selected as photometric binaries (see
figure~\ref{fig:periodcolorexcess}). For the remainder of this
analysis we only consider nonphotometric binaries. The value for $\sin
i$ is determined from:
\begin{equation}
\sin i = \frac{ v \sin i \cdot P}{2 \pi R}
\label{eq:sini}
\end{equation}
where $v \sin i$ is the measured value of the projected equatorial
rotation velocity, $P$ is the measured rotation period, and $R$ is the
stellar radius. We estimate the stellar radius from the $M_{K}$
magnitude using the 125 Myr, solar metallicity (the Pleiades
  have [Fe/H]$=+0.03 \pm 0.05$; \citealp{Soderblom.09}), Yonsei-Yale
isochrone \citep[Y2;][]{Yi.01}. We transform the isochrone to the
2MASS system from the ESO system using the transformations given by
\citet{Carpenter.01}. To determine the absolute magnitudes we assume a
distance to the Pleiades of 133~pc \citep{Soderblom.05} and an
extinction of $A_{K} = 0.01~{\rm mag}$ (S07). Note that if we instead
adopt a distance of 120.2~pc to the cluster based on Hipparcos
parallax measurements \citep{vanLeeuwen.09}, and assume the same
extinction and $M_{K}$-mass and $M_{K}$-radius relations, the inferred
stellar masses and radii at fixed $K_{S}$ magnitude are lower by $\sim
10\%$. Also note that if we use the \citet{Baraffe.98} isochrone
models for stars with $M \sim 0.4~M_{\odot}$, the inferred masses at
fixed $K_{S}$ are lower by $\sim 8\%$, while the inferred radii are
lower by $\sim 30\%$. The inferred radius also depends on the assumed
age for the cluster. The assumed age of $125~{\rm Myr}$ for the
Pleiades from \citet{Stauffer.98} is based on the observed
Li-depletion of main sequence stars in the cluster, a method which has
recently been called into question for other stellar populations
\citep{Jeffries.09, Yee.10}. Determinations of the age of the Pleiades
from the main sequence turn-off range from $\sim 80~{\rm Myr}$
\citep{Mermilliod.81} to $\sim 150~{\rm Myr}$
\citep{Mazzei.89}. Adopting a younger age of $80~{\rm Myr}$ for the
cluster will increase the radii of stars at fixed {\em mass}; the
affect on the $M_{K}$-radius relation, however, is less significant,
as stars with fixed mass will also be brighter in $M_{K}$ at younger
ages. By comparing the $M_{K}$-radius relation obtained from the
$80~{\rm Myr}$ solar-metallicity isochrone from \citet{Siess.00} using
the \citet{Kenyon.95} temperature-colour relations, to the $125~{\rm
  Myr}$ isochrone from the same models\footnote{We use the
  \citet{Siess.00} isochrones rather than the Y2 isochrones because
  the Y2 isochrones do not cover the pre-main sequence stage of
  stellar evolution.}, we estimate that the systematic error in the
radius that may result from overestimating the age of the cluster is
$\la 2\%$ over the magnitude range of interest. Finally if we use the
$M_{V}$ magnitude to determine the radii from the Y2 isochrones,
rather than the $M_{K}$ magnitude, the inferred radii for most stars
are lower by up to $\sim 25\%$.

As seen in Figure~\ref{fig:pervsini}, the $\sin i$ values determined
with equation~\ref{eq:sini} are greater than 1 for many stars with $M
\la 0.85~M_{\odot}$. In Figure~\ref{fig:sinisim} we compare the
observed distribution of $\sin i$ values to various model
distributions. To calculate the model distributions we generate a
sample of $\sin i$ values assuming the rotation axes of the stars are
randomly oriented in space ($\cos i$ is uniformly distributed). We
then assign to each sample $\sin i$ value a gaussian error $\Delta
\sin i$ taken from the set of errors associated with the actual data,
and use these to generate a simulated sample of measured $\sin i_{m}$
values with the condition $\sin i_{m} > 0$. As discussed in
section~\ref{sec:pererr} differential rotation may cause the measured
rotation periods to be systematically longer than the equatorial
rotation periods. This effect will in turn cause the measured $\sin
i_{m}$ values to be systematically greater than the true values. To
include this effect in our models we associate with each simulated
$\sin i$ value a latitude $\beta$ for the dominant spot group. We
assume that these spots may be uniformly distributed over the surfaces
of the stars for reasons outlined in section~\ref{sec:pererr}, so we
draw the values of $\beta$ from a uniform distribution in $\sin
\beta$. We then scale each $\sin i_{m}$ value by the factor $1/(1 - k
\sin^{2}\beta)$ (equation~\ref{eq:diffrot}). Finally we parameterize
any remaining systematic errors in the measured values of $\sin i$ by
multiplying the $\sin i_{m}$ values by a constant scale factor
$\alpha$. We compare the models to the observations using the
Kolmogorov-Smirnov test, recording the probability that the observed
sample is drawn from the same distribution as the simulated sample as
a function of the parameters $k$ and $\alpha$
(figure~\ref{fig:sinisim} panels (c) and (d)). 

For stars with $M > 0.85~M_{\odot}$, we find that the $k = 0$, $\alpha
= 1$ distribution is consistent with the observed distribution (the
K-S probability that the observed and model data-sets are drawn from
the same distribution is $\sim 50\%$), while a $k = 0.08$, $\alpha =
1$ model (i.e. a model with differential rotation that is comparable
to what is expected for young rapidly rotating stars) provides a
slightly better match ($90\%$ probability). A similar result has
recently been found in an independent study of the Pleiades and Alpha
Persei by \citet{Jackson.10} who placed limits on the degree to which
the rotation axes of stars in these clusters are aligned.

For stars with $M < 0.85~M_{\odot}$ the probability that the $k = 0$,
$\alpha = 1$ model is drawn from the same distribution as the
observations is only $\sim 10^{-13}$. In this case a significant value
of $\alpha = 1.24 \pm 0.05$ is required to fit the observations. If we
fix $\alpha = 1.0$ and allow $k$ to vary, the best matching model has
a very high value of $k = 0.60$, with a probability of being drawn
from the same distribution as the observations of less than $30\%$.

While the distribution of $\sin i$ values for $M > 0.85~M_{\odot}$
stars are consistent with the stars having random orientations and
perhaps having a slight degree of differential rotation, the measured
values of $\sin i$ for stars with $M < 0.85~M_{\odot}$ appear to be
systematically too large (the mode is at $\sin i > 1$). The
combination of measured parameters $P v \sin i / R$ may be
systematically larger by a factor of $\sim 1.24$ than the same
combination using the real physical parameters would be. This suggests
systematic errors in one or more of the parameters. Below we consider
each of the parameters in turn.

{\em Radius} -- While it is well-known that the radii of rapidly
rotating stars with $M \la 0.85$ are systematically larger than
theoretical models predict \citep[e.g.][and references
  therein]{Fernandez.09}, the discrepancy appears to be $\sim 10\%$
and not $24\%$. If we adopt the Hipparcos distance of $120.2$\,pc to
the cluster, use the \citet{Baraffe.98} isochrones rather than the Y2
isochrones, or use the $V$ magnitude rather than the $K$ magnitude to
determine the radii, the inferred radii of the stars would be even
smaller, exacerbating the $\sin i$ problem. 

{\em Period} -- While an
extreme differential rotation law of $k = 0.60$ could fit the
distribution, the match is still not very good. Alternatively if the
dominant spot groups on lower mass Pleiades stars are not randomly
distributed, but are instead preferentially located at high latitudes,
this would cause the measured $\sin i$ values to be systematically
larger than they would be if the spots are randomly
distributed. 

{\em $v \sin i$} -- The $v \sin i$ measurements of cool Pleiades
dwarfs could be systematically biased toward larger
values. Figure~\ref{fig:freqpervfreqvsini}a compares the rotation
frequency determined from the photometric period to the rotation
frequency inferred from $v \sin i$ (i.e. $v \sin i / 2\pi R$). For
stars with $M < 0.85~M_{\odot}$, it appears that lower frequency
(longer period) stars are more likely to have $\sin i > 1$ than higher
frequency (shorter period) stars. An offset error in $v \sin i$ would
yield this effect. Most of the longer period, low-mass stars with
$\sin i > 1$ have $v \sin i$ measurements taken from
\citet{Queloz.98}. These authors calculate $v \sin i$ from the
broadening of a spectral cross-correlation function using the
relation:
\begin{equation}
v \sin i = A \sqrt{\sigma^{2} - \sigma_{0}^2}
\end{equation}
where $\sigma^{2}$ is the measured width of the cross-correlation
function and $A$ and $\sigma_{0}^2$ are parameters which are
calibrated by artificially broadening the observed spectra of slowly
rotating field stars to a fixed value of $v \sin i$, and measuring the
widths of the resulting cross-correlation functions. For the ELODIE
spectrograph, \citet{Queloz.98} give $A = 1.9 \pm 0.1$, and
\begin{equation}
\sigma_{0} = 0.27 (B-V)^{2} + 4.51 (\pm 0.06)
\label{eq:sigma0}
\end{equation}
where $B-V$ is the measured color of a star, and $\sigma_{0}$ is
measured in km\,s$^{-1}$. A systematic error in $\sigma_{0}$ would
impact slower rotators more significantly than faster rotators. To estimate the required systematic error in $\sigma_{0}$, we adjust the velocities of stars with $M < 0.85~M_{\odot}$ by
\begin{equation}
\Delta v \sin i = -\frac{A^{2}\sigma_{0}\Delta \sigma_{0}}{v \sin i}
\end{equation}
and use the K-S test to compare the resulting $\sin i$ distribution to
the $\sin i$ distribution for stars with $M > 0.85~M_{\odot}$. For
this test we assume $\sigma_{0} = 4.8$ for these
stars. Figure~\ref{fig:freqpervfreqvsini}c shows the probability that
the two samples are drawn from the same distribution as a function of
$\Delta \sigma_{0}$.  We conducted a similar test assuming a constant
$\Delta v \sin i$ (figure~\ref{fig:freqpervfreqvsini}d).  We find
$\Delta \sigma_{0} = 0.6 \pm 0.2$\,km\,s$^{-1}$ or $\Delta v \sin i =
-1.5 \pm 0.5$\,km\,s$^{-1}$ yields a $\sin i$ distribution for stars
with $M < 0.85~M_{\odot}$ that is statistically indistinguishable from
the distribution for stars with $M > 0.85~M_{\odot}$. If the isochrone
radii of stars with $M < 0.85~M_{\odot}$ are also assumed to be $10\%$
too low, we find $\Delta \sigma_{0} =
0.33^{+0.19}_{-0.27}$\,km\,s$^{-1}$ or $\Delta v \sin i = -0.85 \pm
0.63$.  Figure~\ref{fig:freqpervfreqvsini}b shows how a correction of
$\Delta \sigma_{0} = 0.33$, together with a $10\%$ radius correction
affects the frequencies inferred from $v \sin i$ for stars with $M <
0.85~M_{\odot}$.  A systematic error of $\Delta \sigma_{0} =
0.33^{+0.19}_{-0.27}$\,km\,s$^{-1}$ is consistent at the $1\sigma$
level with the systematic uncertainty in $\sigma_{0}$ of $\sim
0.06$\,km\,s$^{-1}$ estimated by \citet{Queloz.98}. Differences
between the young Pleiades stars and the older field stars used to
calibrate the $(B-V)$-$\sigma_{0}$ relation could result in an even
greater systematic error in $\sigma_{0}$. For example,
\citet{Stauffer.03} showed that Pleiades stars do not have typical
spectral energy distributions--in particular they have excess emission
in the $B$ band (and are therefore {\em bluer} in $B-V$ than expected
from their effective temperatures).  Equation~\ref{eq:sigma0} may
therefore underestimate the intrinsic broadening of Pleiades K
dwarfs. Quantitatively, a $0.1$~mag shift in $B-V$ for Pleiades stars
with $B-V \sim 1.0$ yields an additional $\sim 0.05$\,km\,s$^{-1}$
shift in $\sigma_{0}$.

In conclusion, assuming that the radii are systematically
underestimated by $\sim 10\%$, and that the intrinsic broadening of
these stars are also underestimated by $\sim 0.1$\,km\,s$^{-1}$, we
find that the observed distribution of $\sin i$ values would then be
consistent with the model distribution for a differential rotation
parameter of $k = 0.27 \pm 0.14$ (the $M < 0.85~M_{\odot}$, $\alpha =
1.1$, $\Delta\sigma_{0}=0.1$ line in
figure~\ref{fig:sinisim}d). Allowing for a larger error in the
intrinsic broadening would reduce the required $k$. If we only apply
the radius correction, the differential rotation parameter would need
to be rather large ($k = 0.38^{+0.12}_{-0.15}$; the $M <
0.85~M_{\odot}$, $\alpha = 1.1$ line in figure~\ref{fig:sinisim}d). We
conclude that some combination of the above effects provides a
plausible explanation for the large number of low-mass stars with
measured $\sin i > 1$.

Finally, we note that a qualitatively similar effect has also been
noted by \citet{Jackson.09} for stars with $0.2~M_{\odot} < M <
0.7~M_{\odot}$ in the comparably aged open cluster NGC~2516. They find
that the radii of stars with $M \sim 0.2~M_{\odot}$ must be $\sim
50\%$ larger than theoretical predictions for the $v \sin i$ and
period data to be consistent in this cluster.

\subsection{The Period-Mass Relation and Comparison to Other Clusters}

\begin{figure*}
\includegraphics[width=168mm]{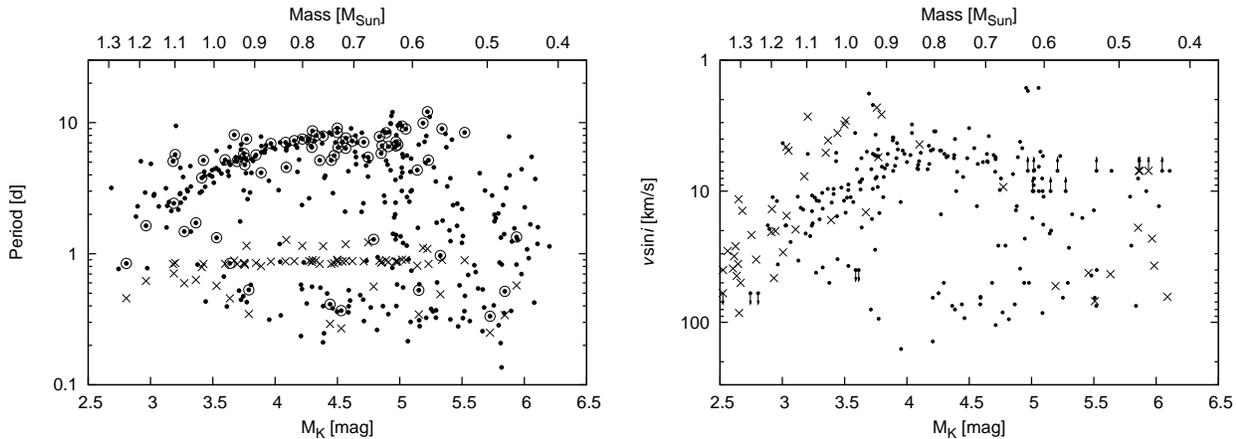}
\caption{Left: photometric rotation period vs. the $K_{S}$ absolute
  magnitude/mass for \NUMperpleiads{} probable members of the Pleiades
  cluster. Note that the period is plotted on a logarithmic
  scale. Filled circles show the adopted period values. Crosses show
  the periods prior to applying the alias correction described in
  section~\ref{sec:alias}. Open circles show the periods after this
  correction. Right: spectroscopically measured $v \sin i$ vs. the
  $K_{S}$ absolute magnitude/mass for stars in our sample with
  measured photometric rotation periods (filled circles) and stars
  without photometric rotation periods (crosses). Note that $v \sin i$
  is plotted on an inverted logarithmic scale. For stars with
  $0.6~M_{\odot} \la M \la 1.0~M_{\odot}$ the sample with measured
  periods is nearly complete, and there does not appear to be any
  significant bias toward selecting fast rotators. For stars with $M
  \ga 1.0~M_{\odot}$ there does appear to be a bias toward detecting
  photometric periods for faster rotators. For stars with $M \la
  0.6~M_{\odot}$ there may be a bias against selecting shorter period
  rotators. This is likely due to rapid rotators having slightly
  redder $(V - K_{S})$ colors at fixed $K_{S}$ magnitude
  (fig.~\ref{fig:periodcolorexcess}), making fast rotators slightly
  fainter in the $r$ filter used by HATNet than slower rotators at
  fixed $K_{S}$.}
\label{fig:permass}
\end{figure*}

\begin{figure}
\includegraphics[width=84mm]{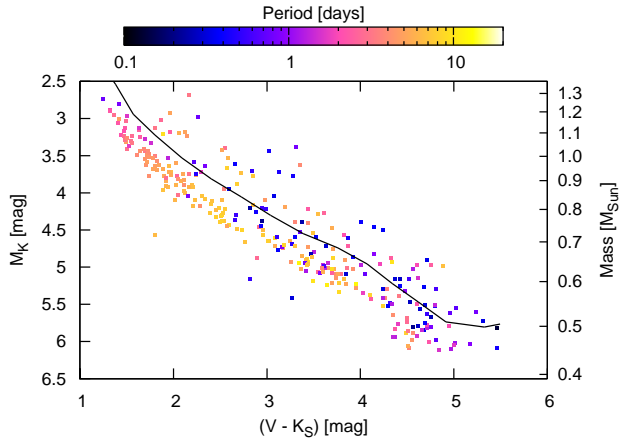}
\caption{$(V-K_{S})$ vs. $M_{K}$ CMD for probable Pleiades members
  with detected rotation periods. The color-intensity of each point is
  scaled by the star's rotation period. At fixed $M_{K}$, short period
  rotators tend to be redder in $(V-K_{S})$ than long period
  rotators. This effect was previously noticed using $v\sin i$ data by
  \citet{Stauffer.03} for Pleiades stars and by \citet{An.07} for
  stars in NGC~2516. \citet{Stauffer.03} suggested that this effect
  may be due to rapid rotators having more red spots than slow
  rotators. Alternatively this may be due to strong magnetic fields in
  rapid rotators inhibiting convection, which causes these stars to be
  larger and cooler than slow rotators
  \citep[e.g.][]{Chabrier.07}. Stars located above the solid line are
  selected as potential photometric binaries, and are excluded from
  the analysis in section~\ref{sec:vsinicomp}.}
\label{fig:periodcolorexcess}
\end{figure}

Figure~\ref{fig:permass} shows the rotation period vs. absolute
magnitude $M_{K}$ and mass, together with the relation between $v \sin
i$ and absolute magnitude $M_{K}$/mass. By comparing the $v \sin i$
values of stars without photometric period detections to the $v \sin
i$ values of stars with period detections, it is apparent that over
the mass range $0.6~M_{\odot} \la M \la 1.0~M_{\odot}$ there does not
appear to be a bias against detecting photometric periods for slow
rotators. For star with $M \ga 1.0~M_{\odot}$ there does appear to be
a bias against detecting periods for the slowest rotators, so our
sample is not complete in this mass range. For stars with $M \la
0.6~M_{\odot}$ the bias appears to be against detecting rapid
rotators. This is likely due to short period stars being redder on
average than long period stars (see Fig.~\ref{fig:periodcolorexcess}
which demonstrates this for $(V-K_{S})$, this is likely to be true for
$(r - K_{S})$ as well), so that at fixed $K_{S}$ short period stars
are fainter in $r$ than long period stars, and therefore have poorer
precision HATNet light curves.

\begin{figure}
\includegraphics[width=84mm]{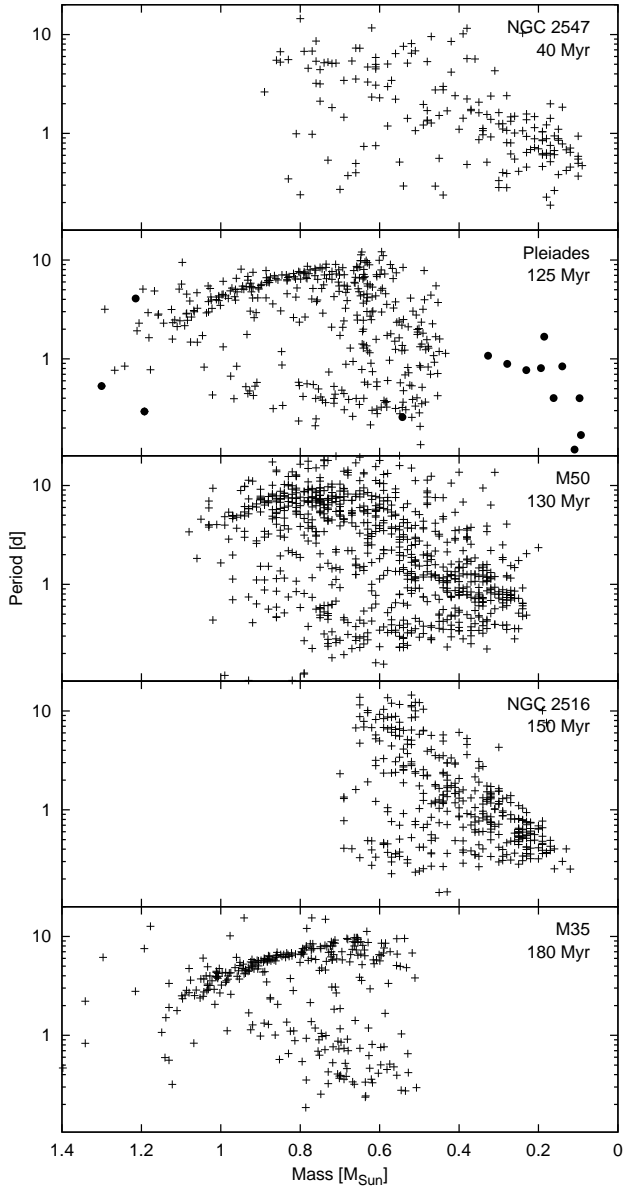}
\caption{Comparison of the mass-period relation in the Pleiades to
  four other open clusters with similar ages. For the Pleiades diagram
  we include 14 stars from the literature which are either too faint
  to be included in the HATNet survey, or did not have periods
  recovered from the HATNet light curves (see
  section~\ref{sec:otherpercomp}). These stars are marked with filled
  circles. The star HD~23386, which has an estimated mass of $\sim
  1.47~M_{\odot}$, falls outside the plotted mass range.}
\label{fig:permassotherclusters}
\end{figure}

In figure~\ref{fig:permassotherclusters} we compare the period-mass
relation for the Pleiades to the relations for four other similarly
aged clusters. These include 3 clusters studied by the MONITOR project
(NGC~2547, \citealp[40~Myr,][]{Irwin.08}; M50,
\citealp[130~Myr,][]{Irwin.09}; and NGC~2516,
\citealp[150~Myr,][]{Irwin.07}), and the cluster M35
\citep[180~Myr,][]{Meibom.09}. For the MONITOR clusters we adopt the
stellar masses given in their tables. For M35 we determine stellar
masses using the extinction corrected $V$-magnitudes given by
\citet{Meibom.09} together with the Y2 isochrones. We assume an age
of 180~Myr \citep{Kalirai.03}, distance of $912~{\rm pc}$
\citep{Kalirai.03}, and metallicity of ${\rm [Fe/H]}=-0.21$
\citep{Schuler.03}.

\begin{figure}
\includegraphics[width=84mm]{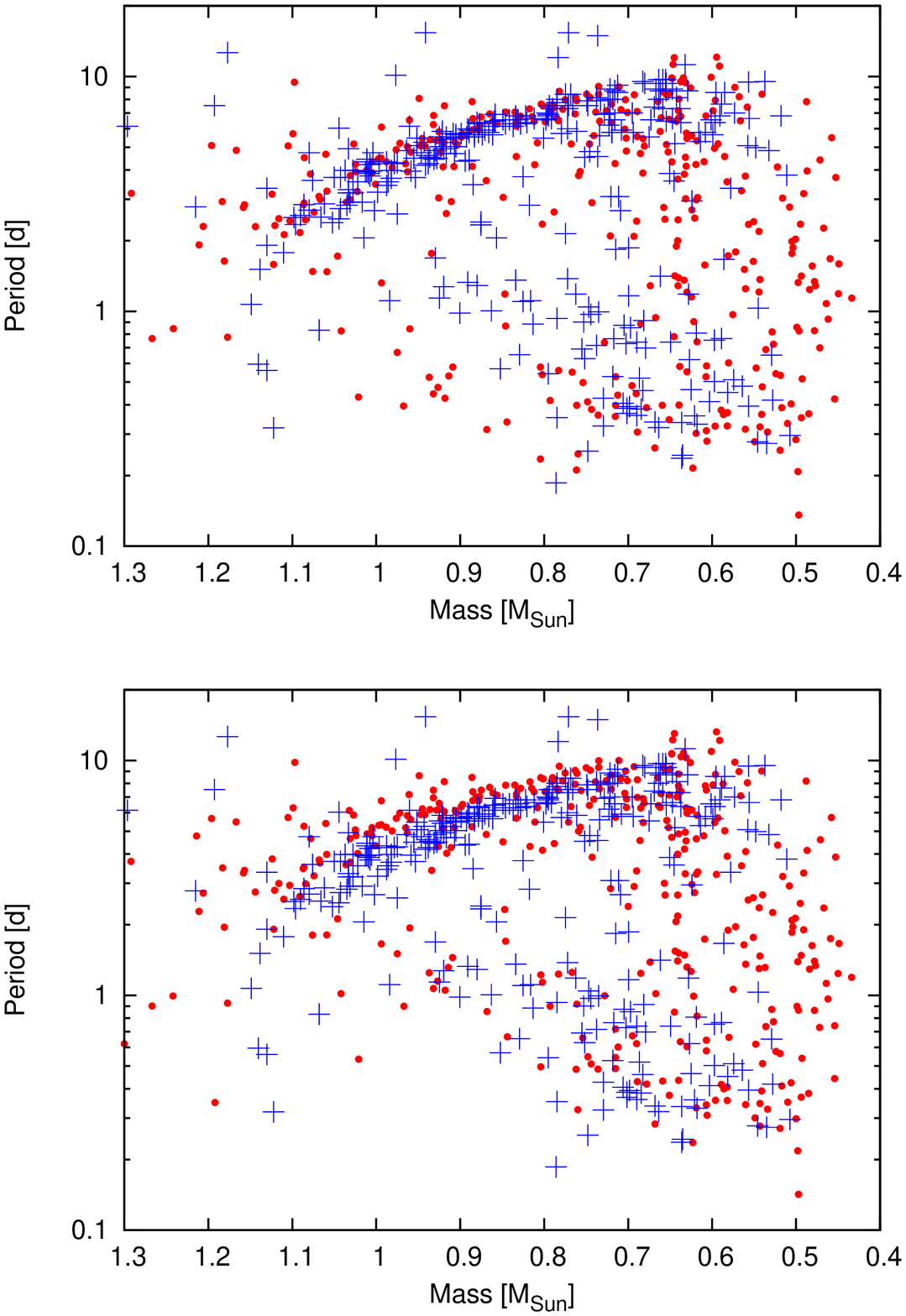}
\caption{Top: Comparison of the mass-period relation in the Pleiades
  (filled circles) and M35 (crosses). Stars on the main
  mass-period sequence with $0.7~M_{\odot} \la M \la 1.1~M_{\odot}$ do
  not show significant period evolution between the $125~{\rm Myr}$
  and $180~{\rm Myr}$ ages of the clusters. Short period rotators in
  this mass range, however, appear to have spun-down. Bottom: Same as
  in the top panel, in this case the Pleiades stars have been evolved
  to the age of M35 using a standard solid-body angular momentum
  evolution model \citep[e.g.][]{Hartman.09a}. We adopt $K_{W} =
  3.19 \times 10^{47}$ for the wind-constant from
  \citet{Denissenkov.09}, and use the mass-dependent values for the
  saturation frequency $\Omega_{\rm sat}$ from
  \citet{Hartman.09a}. Specifically we use $\Omega_{\rm sat} = 5.82
  \Omega_{\odot}$ for $M > 0.99 M_{\odot}$, $\Omega_{\rm sat} = 10.89
  \Omega_{\odot}$ for $0.86 M_{\odot} < M < 0.99 M_{\odot}$,
  $\Omega_{\rm sat} = 8.70 \Omega_{\odot}$ for $0.76 M_{\odot} < M <
  0.86 M_{\odot}$, $\Omega_{\rm sat} = 4.90 \Omega_{\odot}$ for $0.68
  M_{\odot} < M < 0.76 M_{\odot}$, and $\Omega_{\rm sat} = 2.04
  \Omega_{\odot}$ for $M < 0.68 M_{\odot}$. The lack of evolution
  between these two clusters is inconsistent with a solid body
  rotation angular momentum evolution law that is calibrated to the
  rotation of older clusters and the Sun. Models that allow for
  decoupling between the rotation of the core and envelope, on the
  other hand, do predict that the spin-down stalls for the longer
  period stars at $\sim 100~{\rm Myr}$
  \citep{Bouvier.08,Denissenkov.09}.}
\label{fig:compPleiadesM35}
\end{figure}

The data presented here can be combined with data for other open
clusters to test theories of stellar angular momentum evolution. While
a sophisticated analysis like that presented by \citet{Denissenkov.09}
is beyond the scope of this paper, it is instructive to compare the
Pleiades sample to M35. As seen in figure~\ref{fig:compPleiadesM35},
the main period-mass sequences in the two clusters overlap, indicating
that slow rotators with $0.7~M_{\odot} \la M \la 1.1~M_{\odot}$ do not
spin-down between $125~{\rm Myr}$ and $180~{\rm Myr}$. We also show in
figure~\ref{fig:compPleiadesM35} the substantial expected evolution
between the two clusters for a standard solid-body rotation
angular-momentum evolution model \citep[e.g.][]{Hartman.09a}. While
the slowly rotating stars do not spin-down between the ages spanned by
the two clusters, more rapid rotators with $M \ga 0.7~M_{\odot}$ do
appear to have spun-down. These two features can be reproduced by
models that invoke core-envelope decoupling with a coupling time-scale
that depends on the rotation period, such that rapid rotators have a
short coupling time-scale of a few Myr, while slow rotators have a long
time-scale of $\sim 100~{\rm Myr}$
\citep[][]{Bouvier.08,Denissenkov.09}. As noted by \citet{Bouvier.08},
one consequence of slow rotators having a less efficient core-envelope
coupling than rapid rotators is that slower rotators should exhibit
more significant Li depletion than rapid rotators. Such an effect has
been seen by \citet{Soderblom.93} for the Pleiades.

\section{Summary}\label{sec:summary}

In this paper we have presented new rotation period measurements for
\NUMperpleiadsfoundbyhat{} stars in the Pleiades, increasing by a
factor of 5 the number of stars in this cluster with measured
periods. This includes \NUMpleiadsnew{} newly identified probable
cluster members. The sample is $\sim 93\%$ complete for stars with
$0.7~M_{\odot} \la M \la 1.0~M_{\odot}$. By comparing our sample to
the large sample of Pleiades stars with $v \sin i$ measurements in the
literature, we find that our sample is not biased toward short periods
over this same mass range.

We show that for stars with $M \ga 0.85~M_{\odot}$ the inferred
distribution of $\sin i$ values is consistent with the stars having an
isotropic distribution of rotation axes at the $1 \sigma$ level, the
agreement between the model and observations is even better if a $k =
0.08$ differential rotation law is assumed; a differential law of this
form is consistent with theoretical expectations for zero-age main
sequence stars. This result is consistent with the findings of
\citet{Jackson.10}. For stars with $M \la 0.85~M_{\odot}$ the inferred
$\sin i$ values are systematically larger than 1, and is similar to
what was seen by \citet{Jackson.09} for stars in NGC~2516.  Our
observations imply that the combination $P (v \sin i)/R$ is too
large by $\sim 24\%$. We argue that a $\sim 10\%$ systematic error in
the radii of these stars, together with a $\ga 0.1$\,km\,s$^{-1}$
systematic error in the assumed intrinsic broadening of their spectral
lines, and a $k \sim 0.3$ differential rotation law provides a
plausible explanation for this discrepency.

Finally, we find that the mass-period diagram for the Pleiades is
remarkably similar to that seen by \citet{Meibom.09} for the $\sim
185~{\rm Myr}$ cluster M35. In particular there appears to be very
little change in the rotation rates of the slowest rotators with
$0.7~M_{\odot} \la M \la 1.1~M_{\odot}$ between the ages of these
clusters. This result provides support for claims that these stars
have inefficient internal angular momentum transport, and exhibit
significant core-envelope decoupling.

\section*{Acknowledgments}

HATNet operations have been funded by NASA grants NNG04GN74G,
NNX08AF23G and SAO IR\&D grants. G.\'{A}.B. acknowledges support from
the Postdoctoral Fellowship of the NSF Astronomy and Astrophysics
Program (AST-0702843). G.K. thanks the Hungarian Scientific Research
Foundation (OTKA) for support through grant K-81373.  This research
has made use of the SIMBAD database, operated at CDS, Strasbourg,
France. This research has made use of the VizieR catalogue access
tool, CDS, Strasbourg, France. This research has made use of the WEBDA
database, operated at the Institute for Astronomy of the University of
Vienna. This research has made use of the NASA/ IPAC Infrared Science
Archive, which is operated by the Jet Propulsion Laboratory,
California Institute of Technology, under contract with the National
Aeronautics and Space Administration.

\label{lastpage}

\clearpage
\pagestyle{empty}
\begin{landscape}
\begin{table}
\begin{center}{
\scriptsize
\begin{tabular}{lrrlrrrrrrrrrrrrr}
\hline
\multicolumn{1}{c}{ID} & \multicolumn{1}{c}{RAdeg} & \multicolumn{1}{c}{DEdeg} & \multicolumn{1}{c}{OtherID} & \multicolumn{1}{c}{Per} & \multicolumn{1}{c}{ePer} & \multicolumn{1}{c}{fwidth} & \multicolumn{1}{c}{rPer} & \multicolumn{1}{c}{rAmp} & \multicolumn{1}{c}{Bmag} & \multicolumn{1}{c}{Vmag} & \multicolumn{1}{c}{Kmag} & \multicolumn{1}{c}{Mass} & \multicolumn{1}{c}{Radius} & \multicolumn{1}{c}{vsini} & \multicolumn{1}{c}{rvsini} & \multicolumn{1}{c}{fbinary} \\
& \multicolumn{1}{c}{J2000} & \multicolumn{1}{c}{J2000} & & \multicolumn{1}{c}{d} & \multicolumn{1}{c}{d} & \multicolumn{1}{c}{d$^{-1}$} & & \multicolumn{1}{c}{mag} & \multicolumn{1}{c}{mag} & \multicolumn{1}{c}{mag} & \multicolumn{1}{c}{mag} & \multicolumn{1}{c}{$M_{\odot}$} & \multicolumn{1}{c}{$R_{\odot}$} & \multicolumn{1}{c}{km\,s$^{-1}$} & & \\
\hline
HAT-259-0003751 & 51.898273 & 24.528660 &                          DH001 &  7.26243656 &  0.00274337 &  0.00345 & $\cdots$ &  0.0194 & -9.99 & -9.99 &  9.892 & 0.791 & 0.692 &  -99.990000 & $\cdots$ & $\cdots$ \\
HAT-214-0002419 & 52.203186 & 26.499350 &                          DH007 &  6.48431835 &  0.00001564 &  0.00211 & $\cdots$ &  0.0203 & -9.99 & -9.99 &  9.924 & 0.783 & 0.685 &  -99.990000 & $\cdots$ & $\cdots$ \\
HAT-259-0003836 & 52.639614 & 26.215767 &                          DH011 &  4.81593936 &  0.00030922 &  0.00227 & $\cdots$ &  0.0787 & -9.99 & -9.99 & 10.279 & 0.702 & 0.613 &  -99.990000 & $\cdots$ & $\cdots$ \\
HAT-259-0006363 & 52.799591 & 25.165100 &                          DH014 &  8.03366876 &  0.00103605 &  0.00250 & $\cdots$ &  0.0916 & -9.99 & -9.99 & 10.602 & 0.640 & 0.561 &  -99.990000 & $\cdots$ & $\cdots$ \\
HAT-259-0009323 & 52.810955 & 25.981148 &                          DH015 &  2.41116752 &  0.00007308 &  0.00239 & $\cdots$ &  0.0590 & -9.99 & -9.99 & 11.091 & 0.551 & 0.511 &  -99.990000 & $\cdots$ & $\cdots$ \\
HAT-259-0002998 & 52.824719 & 26.028849 & $\cdots$ &  6.37055838 &  0.00006763 &  0.00017 & $\cdots$ &  0.0839 & -9.99 & -9.99 & 10.037 & 0.756 & 0.660 &  -99.990000 & $\cdots$ & $\cdots$ \\
HAT-214-0001101 & 52.890076 & 26.265507 &               PELS008,AKIII170 &  3.24215918 &  0.00016423 &  0.00242 & $\cdots$ &  0.0288 & 11.45 & 10.77 &  9.068 & 1.022 & 0.921 &  15.500000 &  Q98 & $\cdots$ \\
HAT-259-0005281 & 53.001957 & 23.774900 &                        PELS109 &  8.36686239 &  0.00002128 &  0.00204 & $\cdots$ &  0.0601 & 15.27 & 13.95 & 10.520 & 0.655 & 0.573 &  5.500000 &  Q98 & $\cdots$ \\
HAT-259-0004678 & 53.214275 & 25.579544 &                          DH023 &  2.08690706 &  0.00003688 &  0.00256 & $\cdots$ &  0.0848 & -9.99 & -9.99 & 10.323 & 0.693 & 0.605 &  -99.990000 & $\cdots$ & $\cdots$ \\
HAT-259-0003433 & 53.251518 & 25.996056 &                          DH025 &  9.07804569 &  0.00003463 &  0.00215 & $\cdots$ &  0.0264 & -9.99 & -9.99 & 10.128 & 0.735 & 0.642 &  -99.990000 & $\cdots$ & $\cdots$ \\
\end{tabular}
\normalsize
\caption{Catalogue of rotation periods for Pleiades stars. The full table includes 383 stars -- 368 have periods determined from our photometry, and 15 have periods taken from the literature. We exclude the star HII~3167 with a period listed on WEBDA because it is not a cluster member, and we include the star CFHT-PL-8 which is not included in the S07 catalogue. We also include 4 stars excluded from the S07 catalogue, which have previously been identified as candidate cluster members, and which we found to be probable cluster members in section~\ref{sec:newmembers}, and 14 stars which we have newly determined to be probable cluster members. The former stars are SRS~34337, SRS~79807, HCG~84 and HCG~235, while the latter stars can be identified by having no "OtherID" given. The first ID is an internal HAT-ID and is the ID used for labeling the light curves, the other IDs are defined in S07. The formal error on the period from fitting a Gaussian to the peak in the periodogram is given by ePer. The $1\sigma$ width of the Gaussian peak (in frequency) is given by fwidth. Other references with periods for the star in question are given in rPer, the actual periods are given in Table~\ref{tab:knownprd}. The $r$-band peak-to-peak amplitude of the light curve is given by rAmp; stars without an amplitude listed have periods taken from the literature. The mass and radius are estimated from the $K$ magnitude and the YY2 isochrones as described in the text, except for 9 faint stars from \citet{Scholz.04} and 1 faint star from \citet{Terndrup.99} for which the masses and radii are determined using the \citet{Baraffe.98} isochrones. The reference for the $v \sin i$ value is given by rvsini, and fbinary is a flag indicating if the star is a known spectroscopic or visual binary system, it also includes the reference for this determination. Positions and magnitudes are taken from S07. Stars which were not included in the S07 catalogue have positions and magnitudes taken from 2MASS or the 2MASS 6x PSC. For CFHT-PL~8 we take the optical photometry from \citet{Samus.03}, for SRS~34337 and SRS~79807 we take the optical photometry from \citet{Schilbach.95}, while for HAT-259-0000829 we take the optical photometry from the Tycho-2 catalogue \citep{Hog.00}. The full table is included in the supplementary material of the online edition of the journal. A portion is shown here for guidance regarding its form and content. Abreviations for the references are defined in the ReadMe file associated with the electronic version of the table.}
\label{tab:prdcat}
}\end{center}
\end{table}

\begin{table}
\begin{center}{
\scriptsize
\begin{tabular}{lrrlrrrrrrrrrr}
\hline
\multicolumn{1}{c}{ID} & \multicolumn{1}{c}{RAdeg} & \multicolumn{1}{c}{DEdeg} & \multicolumn{1}{c}{OtherID} & \multicolumn{1}{c}{Nobs} & \multicolumn{1}{c}{RMS} & \multicolumn{1}{c}{Bmag} & \multicolumn{1}{c}{Vmag} & \multicolumn{1}{c}{Kmag} & \multicolumn{1}{c}{Mass} & \multicolumn{1}{c}{Radius} & \multicolumn{1}{c}{vsini} & \multicolumn{1}{c}{rvsini} & \multicolumn{1}{c}{fbinary} \\
& \multicolumn{1}{c}{J2000} & \multicolumn{1}{c}{J2000} & & & \multicolumn{1}{c}{mag} & \multicolumn{1}{c}{mag} & \multicolumn{1}{c}{mag} & \multicolumn{1}{c}{mag} & \multicolumn{1}{c}{$M_{\odot}$} & \multicolumn{1}{c}{$R_{\odot}$} & \multicolumn{1}{c}{km\,s$^{-1}$} & & \\
\hline
HAT-259-0000604 & 51.925262 & 23.803688 &                        PELS121 & 6047 &  0.01314 & 10.96 & 10.30 &  8.679 & 1.150 & 1.057 &  4.900000 &  Q98 & $\cdots$ \\
HAT-259-0001639 & 52.168613 & 25.607782 &                        AKIII59 & 6020 &  0.00581 & 12.59 & 11.75 &  9.723 & 0.832 & 0.732 &  -99.990000 & $\cdots$ & $\cdots$ \\
HAT-259-0002068 & 52.409874 & 24.510546 &                          DH008 & 6055 &  0.02236 & -9.99 & -9.99 &  9.698 & 0.839 & 0.738 &  -99.990000 & $\cdots$ & $\cdots$ \\
HAT-259-0001310 & 53.274277 & 22.134232 &          PELS004,DH026,AKII131 & 6046 &  0.00632 & 12.25 & 11.43 &  9.425 & 0.911 & 0.809 &  2.600000 &  Q98 & $\cdots$ \\
HAT-259-0010017 & 53.293736 & 22.522045 &                          DH027 & 5983 &  0.04235 & -9.99 & -9.99 & 11.248 & 0.526 & 0.502 &  -99.990000 & $\cdots$ & $\cdots$ \\
HAT-259-0000347 & 53.530495 & 24.344501 &          PELS006,TrS4,AKIII288 & 6035 &  0.01747 & 10.08 &  9.57 &  8.274 & 1.309 & 1.247 &  35.900000 &  Q98 & $\cdots$ \\
HAT-259-0004453 & 53.684944 & 26.096003 &                          DH037 & 5977 &  0.03269 & -9.99 & -9.99 & 10.371 & 0.683 & 0.597 &  -99.990000 & $\cdots$ & $\cdots$ \\
HAT-259-0000705 & 53.696770 & 26.094885 &               PELS007,AKIII327 & 6009 &  0.00490 & 11.14 & 10.48 &  8.833 & 1.097 & 1.000 &  2.700000 &  Q98 & $\cdots$ \\
HAT-259-0000440 & 53.882030 & 22.823627 &                        PELS124 & 6050 &  0.01401 & 10.40 &  9.86 &  8.541 & 1.200 & 1.115 &  20.500000 &  Q98 & $\cdots$ \\
HAT-259-0001134 & 54.004112 & 24.266083 &                       AKIII391 & 6013 &  0.00495 & 12.06 & 11.29 &  9.388 & 0.922 & 0.820 &  2.300000 &  Q98 & $\cdots$ \\
\end{tabular}
\normalsize
\caption{Catalogue of Pleiades members observed by HATNet without period determinations. The full table includes 120 stars. We include the stars HII~708, HII~727, HII~975, and HHJ~409, which have period measurements taken from the literature, and are included in table~1, but did not have periods recovered from the HATNet light curves. The columns are as in table~1, here we also include Nobs, which is the number of photometric observations, and the unbiased RMS of the HATNet TFA light curve for the target. Only stars with $9.5 < r < 14.5$ are included in this table. The full table is included in the supplementary material of the online edition of the journal. A portion is shown here for guidance regarding its form and content. Abreviations for the references are defined in the ReadMe file associated with the electronic version of the table.}
\label{tab:memnonvarcat}
}\end{center}
\end{table}
\end{landscape}

\clearpage

\pagestyle{empty}
\begin{landscape}
\begin{table}
\begin{center}{
\scriptsize
\begin{tabular}{lrrrrrrr}
\hline
\multicolumn{1}{c}{ID} & \multicolumn{1}{c}{RAdeg} & \multicolumn{1}{c}{DEdeg} & \multicolumn{1}{c}{Per} & \multicolumn{1}{c}{rAmp} & \multicolumn{1}{c}{Jmag} & \multicolumn{1}{c}{Hmag} & \multicolumn{1}{c}{Kmag} \\
& \multicolumn{1}{c}{J2000} & \multicolumn{1}{c}{J2000} & \multicolumn{1}{c}{d} & \multicolumn{1}{c}{mag} & \multicolumn{1}{c}{mag} & \multicolumn{1}{c}{mag} & \multicolumn{1}{c}{mag} \\
\hline
HAT-213-0004432 & 46.376495 & 27.295223 &  1.47490588 &  0.0421 & 11.386 & 11.002 & 10.921 \\
HAT-213-0000301 & 46.395025 & 26.326672 &  0.49278549 &  0.0068 &  7.894 &  7.317 &  7.198 \\
HAT-213-0003310 & 46.423947 & 26.964050 &  0.51638485 &  0.0301 & 10.947 & 10.575 & 10.448 \\
HAT-258-0007599 & 46.452067 & 26.214544 &  0.80443471 &  0.0680 & 11.960 & 11.292 & 11.068 \\
HAT-258-0001512 & 46.545842 & 24.555460 &  1.56383216 &  0.0114 & 10.440 & 10.123 & 10.063 \\
HAT-213-0000501 & 46.549253 & 26.699934 &  1.02287839 &  0.0544 &  7.864 &  6.928 &  6.573 \\
HAT-258-0005170 & 46.567769 & 24.944794 &  7.96319798 &  0.0110 & 11.796 & 11.298 & 11.198 \\
HAT-213-0005056 & 46.574536 & 26.634954 &  6.48431835 &  0.0192 & 11.557 & 11.194 & 11.097 \\
HAT-258-0002836 & 46.593644 & 24.553173 &  0.12926165 &  0.0015 & 11.035 & 10.531 & 10.453 \\
HAT-213-0008723 & 46.597723 & 26.747597 &  0.61090482 &  0.0228 & 12.609 & 12.438 & 12.416 \\
\end{tabular}
\normalsize
\caption{Catalogue of periodic variables in HATNet field G259 that are not cluster members. The full table includes 1804 stars. The positions and magnitudes are taken from 2MASS. The amplitude is the peak-to-peak amplitude of the best-fit sinusoid to the signal-recovery-mode TFA light curve of each star. Note that the true amplitude may be systematically higher by a factor of $\sim 1.3$ (see section~\ref{sec:tfasr}). The full table is included in the supplementary material of the online edition of the journal. A portion is shown here for guidance regarding its form and content.}
\label{tab:fieldprd}
}\end{center}
\end{table}

\begin{table}
\begin{center}{
\scriptsize
\begin{tabular}{lrr}
\hline
\multicolumn{1}{c}{ID} & \multicolumn{1}{c}{Period} & \multicolumn{1}{c}{Ref} \\
& \multicolumn{1}{c}{d} & \\
\hline
BPL106 & 0.170000 & S04 \\
BPL115,DH476 & 0.121250 & S04 \\
BPL125,DH494 & 0.806250 & S04 \\
BPL138,DH505 & 1.075420 & S04 \\
BPL150 & 0.769167 & S04 \\
BPL164,DH544 & 0.840000 & S04 \\
CFHT-PL8 & 0.401000 & T71 \\
HCG20,T1B,DH105 & 2.700000 & K98 \\
HCG254,BPL129 & 0.401667 & S04 \\
HCG346,HHJ111,BPL190 & 1.677920 & S04 \\
\end{tabular}
\normalsize
\caption{Previous rotation period measurements for Pleiades stars. The
  full table includes 94 measurements for 66 stars. The full table is
  included in the supplementary material of the online edition of the
  journal. A portion is shown here for guidance regarding its form and
  content. The references are defined in the ReadMe file associated
  with the electronic version of the table.}
\label{tab:knownprd}
}\end{center}
\end{table}
\end{landscape}


\begin{thebibliography}{}

\bibitem[An et al.(2007)]{An.07} An, D., Terndrup, D.~M., Pinsonneault, M.~H., Paulson, D.~B., Hanson, R.~B., \& Stauffer, J.~R.\ 2007, \apj, 655, 233

\bibitem[Bakos et al.(2004)]{Bakos.04} Bakos, G., Noyes, R.~W., Kov\'{a}cs, G., Stanek, K.~Z., Sasselov, D.~D., \& Domsa, I.\ 2004, \pasp, 116, 266

\bibitem[Bakos et al.(2010)]{Bakos.10} Bakos, G.~\'{A}., et al.\ 2010, \apj, 710, 1724

\bibitem[Baraffe et al.(1998)]{Baraffe.98} Baraffe, I., Chabrier, G., Allard, F., \& Hauschildt, P.~H.\ 1998, \aap, 337, 403

\bibitem[Barnes(2007)]{Barnes.07} Barnes, S.~A.\ 2007, \apj, 669, 1167

\bibitem[Bilir et al.(2008)]{Bilir.08} Bilir, S., Ak, S., Karaali, S., Cabrera-Lavers, A., Chonis, T.~S., \& Gaskell, C.~M.\ 2008, \mnras, 384, 1178

\bibitem[Bouvier et al.(1998)]{Bouvier.98} Bouvier, J., Stauffer, J.~R., Mart\'in, E.~L., Barrado y Navascu\'es, D., Wallace, B., \& B\'ejar, V.~J.~S.\ 1998, \aap, 336, 490

\bibitem[Bouvier(2008)]{Bouvier.08} Bouvier, J.\ 2008, \aap, 489, L53

\bibitem[Brown et al.(2004)]{Brown.04} Brown, B.~P., Browning, M.~K.,
  Brun, A.~S., \& Toomre, J.\ 2004, in Proc. SOHO 14/GONG 2004
  Workshopp (ESA SP-559), Helio- and Asteroseismology: Towards a
  Golden Future, ed. D.~Danesy (Noordwijk: ESA), 341

\bibitem[Carpenter(2001)]{Carpenter.01} Carpenter, J.~M.\ 2001, \aj, 121, 2851

\bibitem[Chabrier et al.(2007)]{Chabrier.07} Chabrier, G., Gallardo, J., \& Baraffe, I.\ 2007, \aap, 472, L17

\bibitem[Clarke, MacDonald, \& Owens(2004)]{Clarke.04} Clarke, D., MacDonald, E.~C., \& Owens, S.\ 2004, \aap, 415, 677

\bibitem[Collier Cameron et al.(2009)]{CollierCameron.09} Collier Cameron, A., et al.\ 2009, \mnras, 400, 451

\bibitem[Deacon \& Hambly(2004)]{Deacon.04} Deacon, N., \& Hambly, N.\ 2004, \aap, 416, 125

\bibitem[Denissenkov et al.(2009)]{Denissenkov.09} Denissenkov, P.~A., Pinsonneault, M., Terndrup, D.~M., \& Newsham, G.\ 2009, \apj submitted, arXiv:0911.1121

\bibitem[Fernandez et al.(2009)]{Fernandez.09} Fernandez, J., et al.\ 2009, \apj, 701, 764

\bibitem[Gray(1976)]{Gray.76} Gray, D.~F.\ 1976, ``The Observation and Analysis of Stellar Photospheres'', Wiley \& Sons Inc (eds.), New York.

\bibitem[Haro et al.(1982)]{Haro.82} Haro, G., Chavira, E., \& Gonzalez, G.\ 1982, Bol. Inst. Tonantzintla, 3, 1

\bibitem[Hartman et al.(2008b)]{Hartman.08b} Hartman, J.~D., Gaudi, B.~S., Holman, M.~J., McLeod, B.~A., Stanek, K.~Z., Barranco, J.~A., Pinsonneault, M.~H., \& Kalirai, J.~S.\ 2008, \apj, 675, 1254

\bibitem[Hartman et al.(2009a)]{Hartman.09a} Hartman, J.~D., et al.\ 2009a, \apj, 691, 342

\bibitem[Hartman et al.(2009b)]{Hartman.09b} Hartman, J.~D., et al.\ 2009b, \aj, submitted, arXiv:0907.2924

\bibitem[H{\o}g et al.(2000)]{Hog.00} H{\o}g, E., et al.\ 2000, \aap, 355, L27

\bibitem[Irwin et al.(2007)]{Irwin.07} Irwin, J., Hodgkin, S., Aigrain, S., Hebb, L., Bouvier, J., Clarke, C., Moraux, E., \& Bramich, D.~M.\ 2007, \mnras, 377, 741

\bibitem[Irwin et al.(2008)]{Irwin.08} Irwin, J., Hodgkin, S., 
Aigrain, S., Bouvier, J., Hebb, L., \& Moraux, E.\ 2008, \mnras, 383, 1588 

\bibitem[Irwin et al.(2009)]{Irwin.09} Irwin, J., Aigrain, S., Bouvier, J., Hebb, L., Hodgkin, S., Irwin, M., \& Moraux, E.\ 2009, \mnras, 392, 1456

\bibitem[Irwin \& Bouvier(2009)]{IrwinBouvier.09} Irwin, J., \& Bouvier, J.\ 2009, IAU Symposium, 258, 363

\bibitem[Jackson et al.(2009)]{Jackson.09} Jackson, R.~J., Jeffries, R.~D., \& Maxted, P.~F.~L.\ 2009, \mnras, 399, L89

\bibitem[Jackson \& Jeffries(2010)]{Jackson.10} Jackson, R.~J., \& Jeffries, R.~D.\ 2010, \mnras, 402, 1380

\bibitem[James et al.(2010)]{James.10} James, D.~J., et al.\ 2010, \aap, in press, arXiv:1004.0047

\bibitem[Jeffries et al.(2009)]{Jeffries.09} Jeffries, R.~D., Jackson, R.~J., James, D.~J., \& Cargile, P.~A.\ 2009, \mnras, 400, 317

\bibitem[Kalirai et al.(2003)]{Kalirai.03} Kalirai, J.~S., Fahlman, G.~G., Richer, H.~B., \& Ventura, P.\ 2003, \aj, 126, 1402

\bibitem[Kawaler(1988)]{Kawaler.88} Kawaler, S.~D.\ 1988, \apj, 333, 236

\bibitem[Kenyon \& Hartmann(1995)]{Kenyon.95} Kenyon, S.~J., \& Hartmann, L.\ 1995, \apjs, 101, 117

\bibitem[Kitchatinov(2005)]{Kitchatinov.05} Kitchatinov, L.~L.\ 2005, Phys.-Usp., 48, 449

\bibitem[K\"onigl(1991)]{Konigl.91} K\"onigl, A.\ 1991, \apj, 333, 236

\bibitem[Kov\'{a}cs, Bakos \& Noyes(2005)]{Kovacs.05} Kov\'{a}cs, G., Bakos, G., \& Noyes, R.~W.\ 2005, \mnras, 356, 557

\bibitem[Krishnamurthi et al.(1998)]{Krishnamurthi.98} Krishnamurthi,
  A., et al.\ 1998, \apj, 493, 914

\bibitem[Lomb(1976)]{Lomb.76} Lomb, N.~R.\ 1976, A\&SS, 39, 447

\bibitem[Mamajek \& Hillenbrand(2008)]{Mamajek.08} Mamajek, E.~E., \& Hillenbrand, L.~A.\ 2008, \apj, 687, 1264

\bibitem[Marilli, Catalano \& Frasca(1997)]{Marilli.97} Marilli, E.,
  Catalano, S., \& Frasca, A.\ 1997, MmSAI, 68, 895

\bibitem[Matt \& Pudritz(2005)]{Matt.05} Matt, S., \& Pudritz, R.~E.\ 2005, \apj, 632, L135

\bibitem[Matt \& Pudritz(2008a)]{Matt.08a} Matt, S., \& Pudritz, R.~E.\ 2008a, \apj, 678, 1109

\bibitem[Matt \& Pudritz(2008b)]{Matt.08b} Matt, S., \& Pudritz, R.~E.\ 2008b, \apj, 681, 391

\bibitem[Mazzei \& Piggato(1989)]{Mazzei.89} Mazzei, P., \& Piggato, L.\ 1989, \aap, 213, L1

\bibitem[Meibom et al.(2009)]{Meibom.09} Meibom, S., Mathieu, R.~D., \& Stassun, K.~G.\ 2009, \apj, 695, 679

\bibitem[Mermilliod(1981)]{Mermilliod.81} Mermilliod, J.~C.\ 1981, \aap, 97, 235

\bibitem[Messina(2001)]{Messina.01a} Messina, S.\ 2001, \aap, 371, 1024

\bibitem[Monet et al.(2003)]{Monet.03} Monet, D.~G., et al.\ 2003, \aj, 125, 984

\bibitem[P\'{a}l \& Bakos(2006)]{Pal.06} P\'{a}l, A., \& Bakos, G.~\'{A}.\ 2006, \pasp, 118, 1474

\bibitem[P\'{a}l(2009)]{Pal.09} P\'{a}l, A.\ 2009, Ph.D. Thesis, E\"otv\"os Lor\'and University, arXiv:0906.3486

\bibitem[Press \& Rybicki(1989)]{Press.89} Press, W.~H., \& Rybicki, G.~B.\ 1989, \apj, 338, 277

\bibitem[Press et al.(1992)]{Press.92} Press, W.~H., Teukolsky, S.~A., Vetterling, W.~T., \& Flannery, B.~P.\ 1992, Numerical Recipes in C, 2nd ed. (New York: Cambridge University Press)

\bibitem[Prosser et al.(1993a)]{Prosser.93a} Prosser, C.~F., Schild,
  R.~E., Stauffer, J.~R., \& Jones, B.~F.\ 1993a, \pasp, 105, 269

\bibitem[Prosser et al.(1993b)]{Prosser.93b} Prosser, C.~F., et
  al.\ 1993b, \pasp, 105, 1407

\bibitem[Prosser et al.(1995)]{Prosser.95} Prosser, C.~F., et al.\ 1995, \pasp, 107, 211

\bibitem[Queloz et al.(1998)]{Queloz.98} Queloz, D., Allain, S., Mermilliod, J.-C., Bouvier, J., \& Mayor, M.\ 1998, \aap, 335, 183

\bibitem[R\"{o}ser et al.(2008)]{Roser.08} R\"{o}ser, S., Schilbach,
  E., Schwan, H., Kharchenko, N.~V., Piskunov, A.~E., \& Scholz,
  R.-D.\ 2008, \aap, 488, 401

\bibitem[Samus et al.(2003)]{Samus.03} Samus, N.~N., et al.\ 2003,
  Astronomy Letters, 29, 468

\bibitem[Scargle(1982)]{Scargle.82} Scargle, J.~D.\ 1982, \apj, 263, 835

\bibitem[Schilbach et al.(1995)]{Schilbach.95} Schilbach, E., Robichon, N., Souchay,
  J., \& Guibert, J.\ 1995, \aap, 299, 696

\bibitem[Scholz \& Eisl\"offel(2004)]{Scholz.04} Scholz, A., \& Eisl\"offel, J.\ 2004, \aap, 421, 259

\bibitem[Schuler et al.(2003)]{Schuler.03} Schuler, S.~C., King, J.~R., Fischer, D.~A., Soderblom, D.~R., \& Jones, B.~F.\ 2003, \aj, 125, 2085

\bibitem[Shu et al.(1994)]{Shu.94} Shu, F., Najita, J., Ostriker, E., Wilkin, F., Ruden, S., \& Lizano, S.\ 1994, \apj, 429, 781

\bibitem[Siess et al.(2000)]{Siess.00} Siess, L., Dufour, E., \& Forestini, M.\ 2000, \aap, 358, 593

\bibitem[Skrutskie et al.(2006)]{Skrutskie.06} Skrutskie, M.~F., et al.\ 2006, \aj, 131, 1163

\bibitem[Skumanich(1972)]{Skumanich.72} Skumanich, A.\ 1972, \apj, 171, 565

\bibitem[Soderblom et al.(1993)]{Soderblom.93} Soderblom, D.~R., Stauffer, J.~R., Hudon, J.~D., \& Jones, B.~F.\ 1993, \apjs, 85, 315

\bibitem[Soderblom et al.(2005)]{Soderblom.05} Soderblom, D., Nelan, E., Benedict, G., McArthur, B., Ramirez, I., Spiesman, W., \& Jones, B.\ 2005, \aj, 129, 161

\bibitem[Soderblom et al.(2009)]{Soderblom.09} Soderblom, D.~R., Laskar, T., Valenti, J.~A., Stauffer, J.~R., \& Rebull, L.~M.\ 2009, \aj, 138, 1292

\bibitem[Stauffer et al.~(1987)]{Stauffer.87} Stauffer, J.~R., Schild,
  R.~A., Baliunas, S.~L., \& Africano, J.~L.\ 1987, \pasp, 99, 471

\bibitem[Stauffer et al.(1998)]{Stauffer.98} Stauffer, J.~R., Schultz, G., \& Kirkpatrick, J.~D.\ 1998, \apj, 499, L199

\bibitem[Stauffer et al.(2003)]{Stauffer.03} Stauffer, J.~R., Burton,
  F.~J., Backman, D., Hartmann, L.~W., Barrado y Navascu\'es, D.,
  Pinsonneault, M.~H., Terndrup, D.~M., \& Muench, A.~A.\ 2003, \aj,
  126, 833

\bibitem[Stauffer et al.~(2007)]{Stauffer.07} Stauffer, J.~R., et al.\ 2007, \apjs, 172, 663

\bibitem[Sukhbold \& Howell(2009)]{Sukhbold.09} Sukhbold, T., \& Howell, S.~B.\ 2009, \pasp, 121, 1188

\bibitem[Terndrup et al.(1999)]{Terndrup.99} Terndrup, D.~M.,
  Krishnamurthi, A., Pinsonneault, M.~H., \& Stauffer, J.~R.\ 1999,
  \aj, 118, 1814

\bibitem[Terndrup et al.(2000)]{Terndrup.00} Terndrup, D.~M., Stauffer, J.~R., Pinsonneault, M.~H., Sills, A., Yuan, Y., Jones, B.~F., Fischer, D., \& Krishnamurthi, A.\ 2000, \aj, 119, 1303

\bibitem[van Leeuwen, Alphenaar, \& Meys(1987)]{VanLeeuwen.87} van
  Leeuwen, F., Alphenaar, P., \& Meys, J.~J.~M.\ 1987, \aaps, 67, 483

\bibitem[van Leeuwen(2009)]{vanLeeuwen.09} van Leeuwen, F.\ 2009, \aap, 497, 209

\bibitem[von Braun et al.(2009)]{vonBraun.09} von Braun, K., et al.\ 2009, IAU Symposium, 253, 478

\bibitem[Walker et al.(2007)]{Walker.07} Walker, G.~A.~H., et al.\ 2007, \apj, 659, 1611

\bibitem[Yee \& Jensen(2010)]{Yee.10} Yee, J.~C., \& Jensen, E.~L.~N.\ 2010, \apj, 711, 303

\bibitem[Yi et al.(2001)]{Yi.01} Yi, S.~K., et al.\ 2001, \apjs, 136, 417

\end{thebibliography}
\end{document}